\documentclass[aps,superscriptaddress,twocolumn,longbibliography]{revtex4-1}
\usepackage{times}
\usepackage{amsmath}
\usepackage{amsfonts}
\usepackage{amssymb}
\usepackage{mathtools}
\usepackage{mathrsfs}
\usepackage{color}
\usepackage[colorlinks=true,citecolor=blue,urlcolor=blue,linkcolor=blue,hyperindex]{hyperref}
\usepackage{wasysym}
\usepackage{bm}
\usepackage{cases}
\usepackage[version=4]{mhchem}
\usepackage{graphicx}
\usepackage{subfigure}
\usepackage[table]{xcolor}
\usepackage{diagbox}
\allowdisplaybreaks

\begin{document}
%%%%%%%%%%%%%%%%%%%%%%%%%%%%%%%%
\title{Level spacing statistics of two unitary equivalent models with distinct local symmetries}
\author{Biquan Yang}
\affiliation{Department of Physics, Chongqing University, Chongqing, 401331, China}
\author{Zijian Xiong}
\email{Corresponding author: xiongzjsysu@hotmail.com}
\affiliation{College of Physics and Electronic Engineering, Chongqing Normal University, Chongqing 401331, China}
\affiliation{Department of Physics, Chongqing University, Chongqing, 401331, China}
\affiliation{Department of Applied Physics, The University of Tokyo, Tokyo 113-8656, Japan}

\begin{abstract}
The full spectrum and integrability of unitary equivalent models are the same. A standard diagnostic tool of integrability is level spacing statistics which requires separating the full spectrum into sectors according to the symmetry. When two unitary equivalent models have different symmetries, it is interesting to know how their level spacing statistics show consistent conclusions. In this work, we examine the level spacing statistics of two unitary equivalent models with distinct local symmetries. The first model is spin-1 XXZ chain $H$, the second model is obtained by Kennedy-Tasaki transformation $U_{KT}HU_{KT}$. We find that the level spacings of model $H$ follow the statistics of the Gaussian orthogonal ensemble (GOE). However, model $U_{KT}HU_{KT}$ only displays GOE statistics in some sectors after a "hidden non-local symmetry" is resolved, and the other sectors labeled by quantum numbers corresponding to the local symmetries exhibit non-GOE statistics. Additionally, a mapping relation between the levels in the sector $\{Z,X,I\}$ of model $H$ and the sector $\{Z,X,I'\}$ of model $U_{KT}HU_{KT}$ is found, where $Z,X,I$ correspond to the quantum numbers of $\pi$ rotation around $z,x$ axes and bond-centered inversion.
\end{abstract}

\maketitle
%%%%%%%%%%%%%%%%%%%%%%%%%%%%%%%%%
\section{Introduction}
Level spacing statistics is a diagnostic tool for distinguishing quantum chaotic and integrable systems \cite{mehta2004,Brody1981rmp,edelman2018chaotic,Guhr1998,Gubin}. For quantum integrable systems, the level spacing distribution typically obeys the Poisson statistics. While it follows the statistics stemming from random matrix theory in the quantum chaotic systems. A defining characteristic of random matrix theory statistics is the level repulsion \cite{wimberger2014}, i.e. the probability of zero level spacing vanishes. In contrast, Poisson statistics exhibit the highest probability around zero level spacing. These two distinct statistics have been confirmed to be universal in various different quantum many-body systems \cite{Montambaux1993prl,Poilblanc1993,Hsu1993prb}, ranging from nuclei systems \cite{Gomez2011}, Hubbard models \cite{Kollath2010,Marco2022prr,Kollath2024} to spin models \cite{Gubin,Santos2004,Santos2020prr,Katsura2023}. 

To obtain the level spacing statistics of a given system, an essential step is to identify the symmetry of the system and separate the eigenvalues into distinct symmetry sectors. Usually, this symmetry refers to the apparent symmetry, like translation, inversion and spin rotation. Recently, it was found that "hidden symmetries" must also be resolved to obtain the correct level spacing statistics \cite{odea2024,Kollath2024}. The consideration of symmetry can be ascribed to the von Neumann-Wigner non-crossing rule \cite{Lieb1971}, which states that the eigenvalues with the same symmetry quantum number rarely cross each other. This implies a phenomenon of level repulsion among eigenvalues within the same symmetry sector. However, level repulsion vanishes if a set of levels data containing levels from different symmetry sectors \cite{Kudo2005,Gubin}. Namely, there will be no level repulsion between levels in different symmetry sectors. Since the level spacing statistics are sensitive to symmetry, some proposals have been made to probe the symmetry \cite{Porter1960pr,Alet2022prx,odea2024}.

For two unitary equivalent models, their full spectrum and integrability are identical. However, in the presence of symmetry, level spacing statistics are only performed on the level spacing data within individual sectors rather than on the entire spectrum. When these two unitary equivalent models possess different symmetries, their full spectrum will be separated in different ways. An intriguing question arises as to how the level spacing statistics of these models yield consistent conclusions. In this work, we consider a non-trivial unitary transformation introduced by Kennedy and Tasaki \cite{Kennedy1992,Kennedy1992prb,Oshikawa1992,tasaki2020}. It is defined in one dimensional integer-spin systems and it reveals that the famous Haldane phase \cite{Haldane1983,Haldane1983prl,ZCGu2009prb,Pollmann2010prb,Pollmann2012prb} can be interpreted as a phase characterized by hidden symmetry breaking \cite{Kennedy1992,Kennedy1992prb,tasaki2020}. It has been pointed out that the transformed Hamiltonian $\tilde{H}$ will have $Z_2\times Z_2$ symmetry which acts locally and short-range interaction, provided that the original Hamiltonian $H$ is invariant under the $\pi$ rotation about the $x,y,z$ axes \cite{Oshikawa1992,Pollmann2012prb} and has only short-range interaction. The symmetry group of $H$ can be higher than $\tilde{H}$. Previous investigations of the Kennedy-Tasaki transformation mainly focus on the low energy properties, such as symmetry protected topological order- spontaneous symmetry breaking duality \cite{li2023non,li2023intrinsically}, and ground state entanglement \cite{Okunishi2011prb}. 

In this paper, we study the level spacing statistics of two unitary equivalent spin-1 models. The original model $H$ is taken to be a spin-1 XXZ chain with single-ion anisotropy which has $U(1)\rtimes Z_2$ internal symmetry. The transformed Hamiltonian $\tilde{H}$ is a model with $Z_2\times Z_2$ internal symmetry and two-body short-range interactions. We find that the level spacings of model $H$ follow GOE statistics while the transformed model $\tilde{H}$ do not. Only after a "hidden symmetry" is taken into account, certain sectors of model $\tilde{H}$ exhibit GOE statistics, but the remaining sectors still show non-GOE statistics. The paper is organized as follows. In Sec.\ref{sectmod}, we describe the models studied in this paper and their symmetries. In Sec.\ref{sectlss} and Sec.\ref{sectres}, the level spacing statistics methods and results of the models are presented. Conclusions are provided in Sec.\ref{sectcon}.

%%%%%%%%%%%%%%%%%%%%%%%%%%%%%%%%%%%
\section{Models}\label{sectmod}
The spin-1 XXZ chain with single-ion anisotropy under the open boundary condition is given by
\begin{equation}\label{hal}
H=\sum_{i=1}^{L-1}(S^{x}_{i}S^{x}_{i+1}+S^{y}_{i}S^{y}_{i+1}+\Delta S^{z}_{i}S^{z}_{i+1})+D\sum_{i=1}^{L}(S_{i}^{z})^2.
\end{equation}
This model has been well studied and exhibits a rich phase diagram including the Haldane phase and several symmetry-breaking phases \cite{WeiChen2003prb}. 

The internal symmetry of this model is $U(1)\rtimes Z_2$, where the $U(1)$ is spin rotation around the $z$ axis: $U_{\theta}^{z}=e^{-i\theta \sum_{i=1}^{L}S_i^z}$. And the $Z_2$ is the spin flip, which is equivalent to the $\pi$ rotation around the $x$ axis (denoted by $U^{x}_\pi=e^{-i\pi \sum_{i=1}^{L}S_i^x}$) or $y$ axis. By utilizing the identity $(S^{\alpha})^3=S^{\alpha}$ for spin-1 (where $\alpha=x,y,z$), one can find the $\pi$ rotation for a single spin has a simple form \cite{tasaki2020}: $U^{\alpha}_{\pi}=1-2(S^{\alpha})^2$. This model also has a spatial symmetry which is bond-centered inversion symmetry $\mathcal{I}: S_{i}^{\alpha}\to S_{L-i+1}^{\alpha}$. 

In this paper, we mainly focus on a model $\tilde{H}$ which is unitary equivalent to $H$ in eq.(\ref{hal})
\begin{equation}\label{dualhal}
\begin{aligned}
\tilde{H}=&U_{KT}HU_{KT}^{\dagger}\\
=&\sum_{i=1}^{L-1}[-S^{x}_{i}S^{x}_{i+1}+S^{y}_{i}exp(i\pi S^{z}_{i})exp(i\pi S^{x}_{i+1})S^{y}_{i+1}\\
&-\Delta S^{z}_{i}S^{z}_{i+1}]+D\sum_{i=1}^{L}(S_{i}^{z})^2,
\end{aligned}
\end{equation}
where $U_{KT}$ is the Kennedy-Tasaki transformation \cite{Kennedy1992,Kennedy1992prb,Oshikawa1992}, it is a non-local unitary transformation defined on the chain with open boundary condition
\begin{equation}
U_{KT}=\prod_{1\leq u<v \leq L}e^{i\pi S^{z}_{u}S^{x}_{v}}.
\end{equation}
Based on the algebra for spin-1, one can find that \cite{tasaki2020} $(U_{KT})^2=1$ and $U_{KT}=(U_{KT})^{\dagger}$. Under the Kennedy-Tasaki transformation, the spin operators are transformed as \cite{Kennedy1992,Kennedy1992prb,Oshikawa1992}
\begin{equation}
\begin{aligned}\label{KTs}
U_{KT}S_{j}^{x}U_{KT}=&S_{j}^{x}exp(i\pi\sum_{k=j+1}^{L}S^{x}_{k}),\\
U_{KT}S_{j}^{z}U_{KT}=&exp(i\pi\sum_{k=1}^{j-1}S^{z}_{k})S_{j}^{z},\\
U_{KT}S_{j}^{y}U_{KT}=&exp(i\pi\sum_{k=1}^{j-1}S^{z}_{k})\,S_{j}^{y}\,exp(i\pi\sum_{m=j+1}^{L}S^{x}_{m}),
\end{aligned}
\end{equation}
where the transformed operators become complicated and non-local. Although the transformation is non-local, it is clear that \emph{model $\tilde{H}$ is still a simple model with only two-body short-range interaction}.

The only local internal symmetry of model $\tilde{H}$ which acts on-site \cite{Kennedy1992,Kennedy1992prb,Oshikawa1992} is $Z_{2}\times Z_{2}$, it is generated by the $\pi$ rotation around the $x,y,z$ axes. Here the local symmetry means the symmetry transformation operator can be written as a product of unitary operators acting on each site \cite{tasaki2020}. And this symmetry is the subgroup of $U(1)\rtimes Z_2$ in model $H$. This can be seen by $U_{KT}(S^{\alpha}_{i})^{2}U_{KT}=(S^{\alpha}_{i})^{2}$, which implies the $\pi$ rotation is invariant under the transformation: $U_{KT}U^{\alpha}_{\pi}U_{KT}=U^{\alpha}_{\pi}$. It is well known that the symmetry breaking of this $Z_2\times Z_2$ explains the hidden string order in Haldane phase and also the degeneracy of the ground state in the spin-1 chain with open boundary condition \cite{tasaki2020,Kennedy1992,Kennedy1992prb,Oshikawa1992}. Model $\tilde{H}$ also has a "hidden non-local symmetry", the corresponding generator is $U_{KT}\sum_{i}S^{z}_{i}U_{KT}$. Later we will study the effect of this "hidden symmetry".

Consider the bond-centered inversion, 
\begin{equation}
\begin{aligned}
&\mathcal{I}: S^{y}_{i}exp(i\pi S^{z}_{i})exp(i\pi S^{x}_{i+1})S^{y}_{i+1}\\
 \to& S^{y}_{L-i+1}exp(i\pi S^{z}_{L-i+1})exp(i\pi S^{x}_{L-i})S^{y}_{L-i}\\
 =&exp(i\pi S^{x}_{L-i})S^{y}_{L-i}S^{y}_{L-i+1}exp(i\pi S^{z}_{L-i+1}).
\end{aligned}
\end{equation}
Using the matrix representation of the spin-1 operators \cite{tasaki2020}, one can find
 \begin{equation}\label{spcon}
 S^{y}_{i}exp(i\pi S^{z}_{i})=exp(i\pi S^{x}_{i})S^{y}_{i},
 \end{equation}
Thus, $\sum_{i}S^{y}_{i}exp(i\pi S^{z}_{i})exp(i\pi S^{x}_{i+1})S^{y}_{i+1}$ is invariant under the bond-centered inversion, and model $\tilde{H}$ is also invariant under this symmetry. 

Model $H$ and model $\tilde{H}$ are unitary equivalent to each other, thus their full spectrum are identical. Nonetheless, their apparent symmetries are different, and we will see this leads to different level spacing statistics in their respective sectors.

%%%%%%%%%%%%%%%%%%%%%%%%%%%%%%%%%%%
\section{Level spacing statistics}\label{sectlss}
In this paper, the eigenvalues of the models are obtained by numerical diagonalization \cite{quspin1,quspin2}. We denote a set of eigenvalues of the Hamiltonian as $\{E_{n}\}_g$, where $g$ labels a sector with fixed quantum numbers. We denote the eigenvalues of $\sum_{i=1}^{L}S^{z}_{i}$, $U^{x}_{\pi}$, $U^{z}_{\pi}$ and $\mathcal{I}$ as $M^{z}, X, Z, I$, respectively. For model $H$, $g$ represents a set $\{M^z, I\}$ for $M^z\neq0$ and $\{M^z=0, X,I\}$. For model $\tilde{H}$, $g$ is $\{Z,X,I\}$. The eigenvalues are sorted in ascending order, $E_1\leq E_2\leq \cdots$. Level spacing is $\delta_n=E_{n+1}-E_n$. 

We use the $S^z$ direct product basis to block diagonalize the model $H$. The eigenstates of $S^z$ are $|1\rangle, |0\rangle, |-1\rangle$. To block diagonalize the model $\tilde{H}$, we can use the "$Z_2\times Z_2$" symmetry basis \cite{Kennedy1994,HongYang2023prb}
\begin{equation}
\begin{aligned}\label{z2basis}
|\uparrow\rangle=&\frac{1}{\sqrt{2}}(|1\rangle-|-1\rangle),\\
|\downarrow\rangle=&|0\rangle,\\
|h\rangle=&\frac{1}{\sqrt{2}}(|1\rangle+|-1\rangle),
\end{aligned}
\end{equation}
which are the simultaneous eigenstates of $U_{\pi}^{x},U_{\pi}^{y},U_{\pi}^{z}$,
\begin{equation}
\begin{aligned}\label{zxeigen}
&U^{x}_{\pi}|\uparrow\rangle=|\uparrow\rangle,\,U^{x}_{\pi}|\downarrow\rangle=-|\downarrow\rangle,\,U^{x}_{\pi}|h\rangle=-|h\rangle,\\
&U^{y}_{\pi}|\uparrow\rangle=-|\uparrow\rangle,\,U^{y}_{\pi}|\downarrow\rangle=-|\downarrow\rangle,\,U^{y}_{\pi}|h\rangle=|h\rangle,\\
&U^{z}_{\pi}|\uparrow\rangle=-|\uparrow\rangle,\,U^{z}_{\pi}|\downarrow\rangle=|\downarrow\rangle,\,U^{z}_{\pi}|h\rangle=-|h\rangle.
\end{aligned}
\end{equation}

There are several quantities for studying the level spacing statistics. The first one is based on the "normalized" level spacing $\tilde{\delta}$ rather than the raw level spacing \cite{Gubin,edelman2018chaotic}. The "normalization" procedure is called unfolding, which sets the mean level density to 1. There are different unfolding approaches \cite{Gubin,Kudo2003prb}. For an integrable system, the distribution of "normalized" level spacings follows Poisson statistics $P_{P}(\tilde{\delta})=e^{-\tilde{\delta}}$. The chaotic system with time-reversal symmetry or the real symmetric Hamiltonian (the models in this paper all belong to this class) can be described by random matrices from the Gaussian orthogonal ensemble. And the distribution of "normalized" level spacings has the form $P_{GOE}(\tilde{\delta})=\frac{\pi\tilde{\delta}}{2}exp(-\pi\tilde{\delta}^2/4)$, which reflects the level repulsion by $P_{GOE}(\tilde{\delta}\to 0)\to 0$.

An alternative quantity based on the gap ratio between consecutive levels is constructed in ref.\cite{Huse2007prb}, 
\begin{equation}
r_{n}=\frac{\rm{min}(\delta_n,\delta_{n+1})}{\rm{max}(\delta_{n},\delta_{n+1})}.
\end{equation}
The unfolding procedure is not needed for the computation of this quantity \cite{Huse2007prb,Atas2013prl}. The gap ratio distribution $P(r)$ is $P_{P}(r)=2/(1+r)^2$ for Poisson statistics and $P_{GOE}(r)=\frac{27}{4}\frac{r+r^2}{(1+r+r^2)^{5/2}}$ for GOE statistics \cite{Atas2013prl}. The mean gap ratio $\langle r\rangle$ is also widely used \cite{Huse2007prb,Atas2013prl}. It is found that $\langle r\rangle_P\approx0.386$ for Poisson statistics, $\langle r\rangle_{GOE}\approx0.536$ for GOE. In the following, we use both the gap ratio distribution and mean gap ratio as chaotic indicators to study the spectrum.

 %%%%%%%%%%%%%%%%%%%%%%%%%
\section{results}\label{sectres}
\subsection{Different level spacing statistics in model $H$ and model $\tilde{H}$}
\begin{figure}
	\centering
	\includegraphics[width=0.5\textwidth]{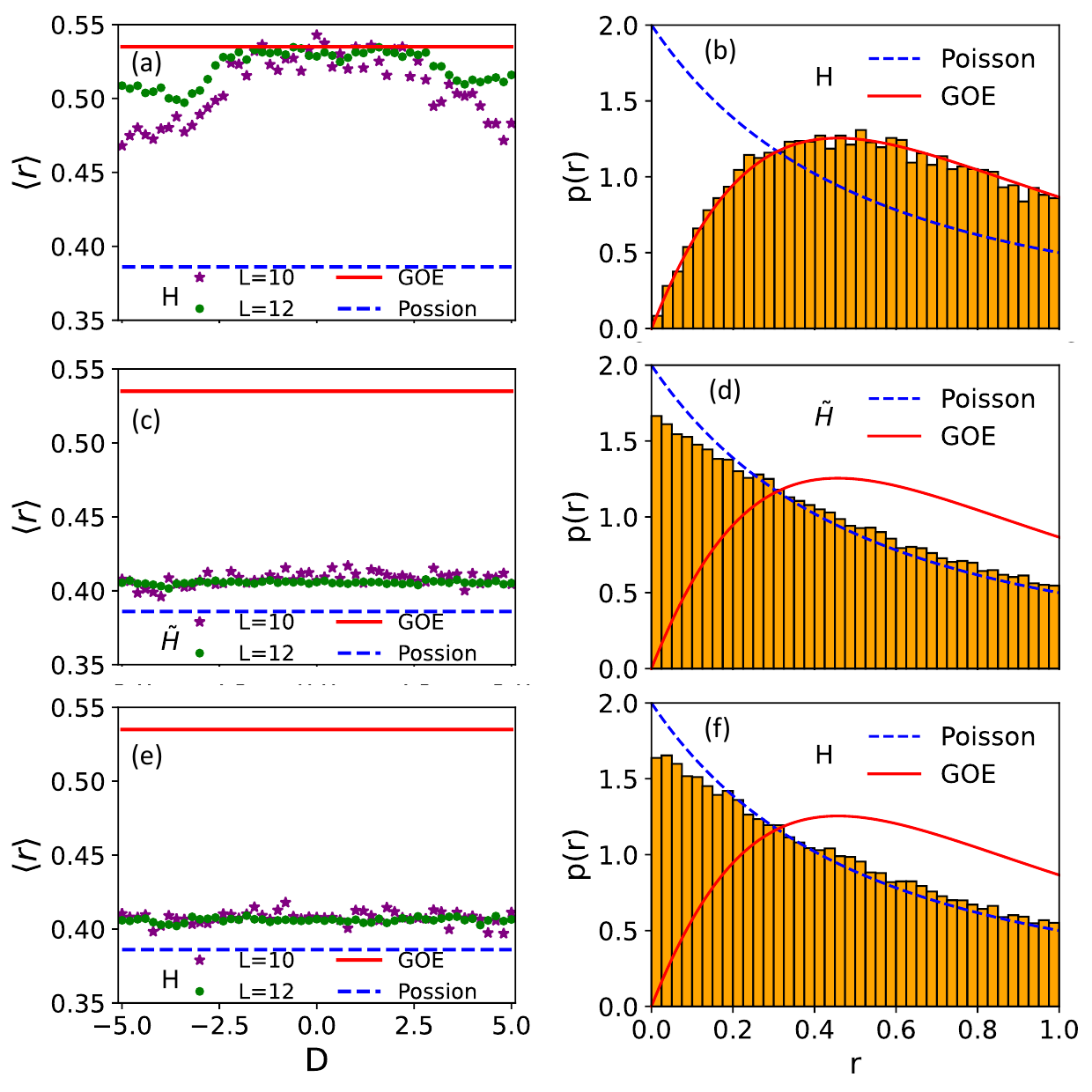}
	\caption{\label{fig1} Mean gap ratio $\langle r\rangle$ as a function of single-ion anisotropy $D$ for (a) model $H$ in sector $\{M^z=0,X=-1,I=1\}$, (e) in sector $\{Z=1,X=-1,I=1\}$, and for (c) model $\tilde{H}$ in sector $\{Z=1,X=-1,I=1\}$. Gap ratio distribution $P(r)$ with $D=1.2$ and system size $L=12$ for (b) model $H$ in sector $\{M^z=0,X=-1,I=1\}$, (f) in sector $\{Z=1,X=-1,I=1\}$, and for (d) model $\tilde{H}$ in sector $\{Z=1,X=-1,I=1\}$. The red (solid) line and the blue (dashed) line represent the prediction from GOE and Poisson statistics, respectively.}
\end{figure}

Though the random matrix theory statistics have been found in the distribution of level spacings of certain related spin-1 models, such as the Affleck-Kennedy-Lieb-Tasaki model and its variants \cite{Sanjay2018prb,Katsura2023}. To the best of our knowledge, the level spacing statistics of the model $H$ have not been reported, it is expected to follow GOE distribution. Without loss of generality, we fixed $\Delta=0.5$ in this paper. In Fig.\ref{fig1}(a), we show the dependence of the mean gap ratio on anisotropy $D$ for the model $H$ in sector $\{M^z=0, X=-1, I=1\}$. This sector has dimensions 2190, and 18300 for system size $L=10, 12$, respectively. We find a good agreement between the mean gap ratio and the prediction from the GOE statistics ($\langle r\rangle_{GOE}\approx0.536$) for $-2<D<2$. But the mean gap ratio deviates from the GOE value for the larger parameter (here, $|D|$), which is also found in other models \cite{Kollath2010,Marco2022prr,Kollath2024}. This deviation shows an obvious system size dependence. For larger system sizes, the deviation decreases and the mean gap ratio moves toward the GOE value. In Fig.\ref{fig1}(b), we show the gap ratio distribution of model $H$. It is clear that it follows GOE prediction nicely, and we have also checked other parameters and sectors with large enough dimensions, they all show GOE statistics. Hence, we can conclude that model $H$ is chaotic.

\begin{figure}
	\centering
	\includegraphics[width=0.5\textwidth]{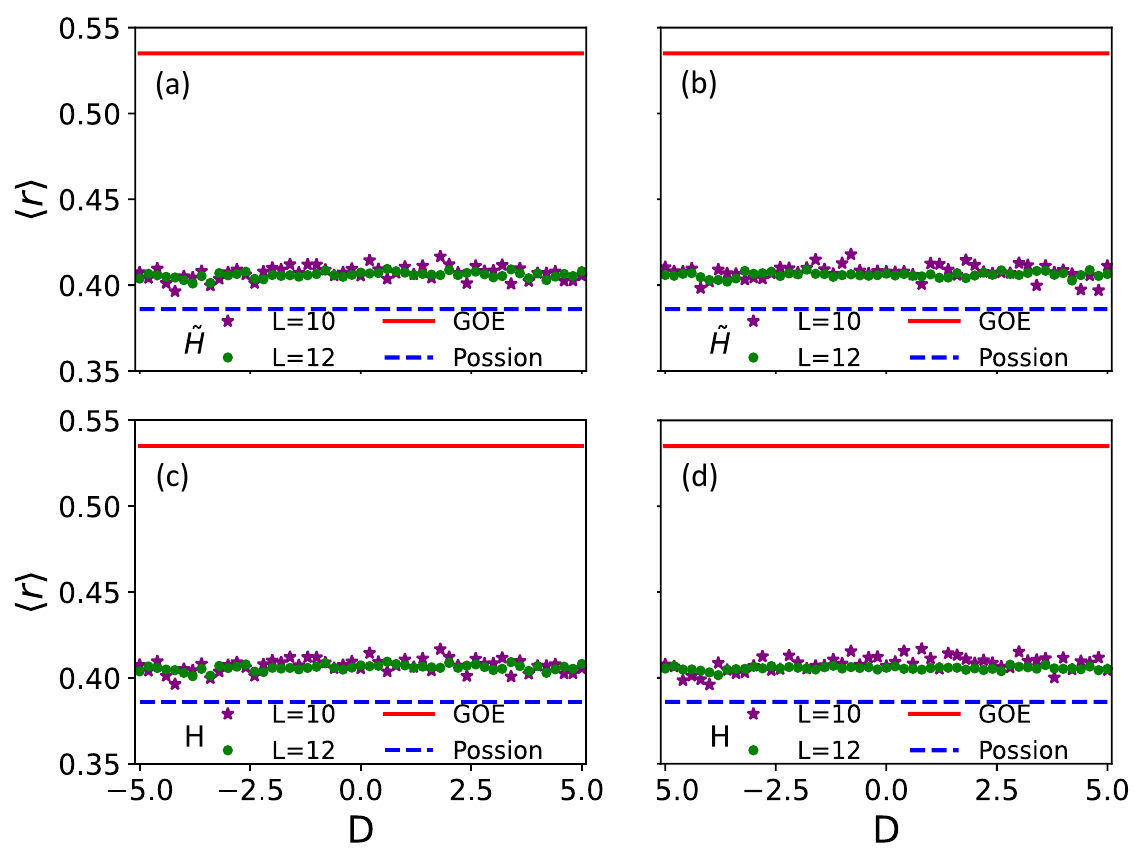}
	\caption{\label{fig2} Mean gap ratio $\langle r\rangle$ as a function of single-ion anisotropy $D$ for (a) model $\tilde{H}$ in sector $\{Z=1,X=1,I=1\}$, (b) in sector $\{Z=1,X=-1,I=-1\}$, and for (c) model $H$ in sector $\{Z=1,X=1,I=1\}$, (d) in sector $\{Z=1,X=-1,I=-1\}$. The red (solid) line and the blue (dashed) line represent the prediction from GOE and Poisson statistics, respectively.}
\end{figure}

As model $\tilde{H}$ is unitary equivalent to model $H$, there is no doubt that model $\tilde{H}$ should also be chaotic. But these two models have different apparent symmetries, level spacing statistics should be done in their respective apparent symmetry sectors. For model $\tilde{H}$, the sectors are labeled by sets $\{Z,X,I\}$. In Fig.\ref{fig1}(c), we show the dependence of the mean gap ratio on anisotropy $D$ for model $\tilde{H}$ in sector $\{Z=1, X=-1, I=1\}$. This sector has dimensions 7381, 66430 for system size $L=10, 12$, respectively. Interestingly, the mean gap ratio of model $\tilde{H}$ follows neither GOE statistics nor Poisson statistics. The gap ratio distribution in this sector is shown in Fig.\ref{fig1}(d), even though there is no level repulsion, it does not follow the Poisson statistics. Other sectors also show similar non-GOE statistics, see Fig.\ref{fig2} (a) (b).

The deviation from random matrix theory statistics observed in the $\{Z,X,I\}$ sector of model $\tilde{H}$ can be understood as follows. As the $\pi$ rotations are invariant under the Kennedy-Tasaki transformation, the quantum numbers $Z,X$ remain unchanged under this unitary transformation. Consequently, the corresponding spectrum of $\{Z,X\}$ sector of model ${H}$ is the same as in the $\{Z,X\}$ sector of model $\tilde{H}$. On the other hand, every $\{Z,X\}$ sector of model $H$ contains levels from different $\{M^z\}$ sectors. For example, all levels within the sector $\{M^z=0,X=-1\}$ are included in the sector $\{Z=1,X=-1\}$, and $\{M^{z}\neq0\}$ sectors with $M^z$ even are also included. It is well known that a spectrum coming from different symmetry sectors will shows Poisson-like statistics \cite{Alet2022prx}, thus $\{Z,X,I\}$ sectors of model $H$ do not follow GOE statistics. In Fig.\ref{fig1}(e)(f), we show the mean gap ratio and gap ratio distribution in the sector $\{Z=1,X=-1,I=1\}$ of model ${H}$, which show non-GOE behavior. Accordingly, the $\{Z,X,I\}$ sectors of model $\tilde{H}$ also show non-GOE statistics since the spectrum is the same as model $H$ in the corresponding sectors. Note that the number of $M^z\neq0$ sectors grows with the system size, more symmetry sectors will be mixed for larger systems and such non-GOE behavior may evolve to Poisson statistics in the thermodynamic limit.

%%%%%%%%%%%%%%%%%%%%%%
\subsection{Bond-centered inversion and the Kennedy-Tasaki transformation}
However, a careful comparison between Fig.\ref{fig1}(c) and (e) shows the spectrum of model $\tilde{H}$ is different from the model $H$ in $\{Z=1,X=-1,I=1\}$ sector. More sectors are shown in Fig.\ref{fig2}. In Fig.\ref{fig2}(a)(c), we find that model $H$ and model $\tilde{H}$ share identical spectra in the sector $\{Z=1,X=1,I=1\}$. This sector has dimensions 7503, 66795 for system size $L=10,12$, respectively. While the spectrum of model $\tilde{H}$ in the sector $\{Z=1, X=-1,I=-1\}$ (Fig.\ref{fig2}(b)) differs from that of model $H$ in the same sector (Fig.\ref{fig2}(d)), but it is the same as the spectrum of model $H$ in the sector $\{Z=1,X=-1,I=1\}$ (Fig.\ref{fig1}(e)). Furthermore, the spectrum of model $\tilde{H}$ in the sector $\{Z=1,X=-1,I=1\}$ (Fig.\ref{fig1}(c)) is the same as the model $H$ in the sector $\{Z=1,X=-1,I=-1\}$ (Fig.\ref{fig2}(d)). $\{Z=1,X=-1,I=-1\}$ sector has dimensions 7381, 66430 for system size $L=10,12$, respectively. A summary of these findings is given in Table.\ref{tabsum}. These results show that the spectrum of these two models may not coincide in every $\{Z,X,I\}$ sector, while the full spectrum of them are the same in every $\{Z,X\}$ sector. And these findings also suggest a potential relationship between the spectrum of model $\tilde{H}$ in sector $\{Z,X,I\}$ and model $H$ in sector $\{Z,X,I'\}$.

\begin{table}
\caption{Summary of the level spacing statistics of model $\tilde{H}$ and model $H$ in $\{Z,X,I\}$ sectors in Fig.\ref{fig1} and Fig.\ref{fig2}. The same shaded colors (font styles) in the second and third columns mean the spectrum of the corresponding sectors is the same.}
\begin{ruledtabular}
\label{tabsum}
\begin{tabular}{c|l|l}
\diagbox{Sectors}{Models} & $\tilde{H}$ & $H$ \\
\hline
$\{Z=1,X=-1,I=1\}$ & \cellcolor{gray!30}Fig.\ref{fig1} (c) & \cellcolor{blue!30}\textbf{Fig.\ref{fig1} (e)}\\
\hline
$\{Z=1,X=1,I=1\}$ & \cellcolor{green!30}\textit{Fig.\ref{fig2} (a)} & \cellcolor{green!30}\textit{Fig.\ref{fig2} (c)}\\
\hline
$\{Z=1,X=-1,I=-1\}$ & \cellcolor{blue!30}\textbf{Fig.\ref{fig2} (b)} & \cellcolor{gray!30}Fig.\ref{fig2} (d)
\end{tabular}
\end{ruledtabular}
\end{table}

Since the quantum numbers $Z,X$ are invariant under the Kennedy-Tasaki transformation, to formulate the relation between the spectrum in model $H$ and model $\tilde{H}$, we need to consider the bond-centered inversion symmetry under the transformation. Before this, we need to introduce some setup. Suppose $|\psi\rangle$ is a simultaneous eigenstate of $H$, $U^{z}_{\pi}$, $U^{x}_{\pi}$ and $\mathcal{I}$, i.e. $H|\psi\rangle=E|\psi\rangle$, $U^{z}_{\pi}|\psi\rangle=Z|\psi\rangle$, $U^{x}_{\pi}|\psi\rangle=X|\psi\rangle$, $\mathcal{I}|\psi\rangle=I|\psi\rangle$. Then, $U_{KT}|\psi\rangle$ is an eigenstate of $\tilde{H}$, that is $\tilde{H}U_{KT}|\psi\rangle=EU_{KT}|\psi\rangle$. Note that $U_{KT}$ and $\mathcal{I}$ do not commute, 
\begin{equation}
\begin{aligned}
&\mathcal{I}U_{KT}\mathcal{I}^{-1}\\
=&exp(i\pi S^{z}_{L}S^{x}_{L-1})exp(i\pi S^{z}_{L}S^{x}_{L-2})\cdots exp(i\pi S^{z}_{L}S^{x}_{1})\\
&exp(i\pi S^{z}_{L-1}S^{x}_{L-2})exp(i\pi S^{z}_{L-1}S^{x}_{L-3})\cdots exp(i\pi S^{z}_{L-1}S^{x}_{1})\\
&\cdots\\
&exp(i\pi S^{z}_{2}S^{x}_{1})\\
=&\prod_{1\leq v<u \leq L}e^{i\pi S^{x}_{v}S^{z}_{u}}\equiv U_{KT}^{>},
\end{aligned}
\end{equation}
thus, $[U_{KT},\mathcal{I}]\neq 0$. It is not difficult to verify that $[U_{KT}, U_{KT}^{>}]=0$, $(U_{KT}^{>})^2=1$ and the quantum numbers $Z,X$ are also invariant under $U_{KT}^{>}$. 

Consider a direct product state $|\alpha\rangle$ in the $Z_2\times Z_2$ basis, e.g. $|\alpha\rangle=|\uparrow\uparrow h\cdots\rangle$. Suppose the number of $\uparrow$, $\downarrow$, $h$ in $|\alpha\rangle$ are $n_{u},n_{d},n_{h}$. Provided the system size is $L$, then $n_{u}+n_{d}+n_{h}=L$. Using eq.(\ref{zxeigen}), we have
\begin{equation}
\begin{aligned}\label{uktalpha}
U^{x}_{\pi}|\alpha\rangle=&(-1)^{n_d+n_h}|\alpha\rangle,\\
U^{z}_{\pi}|\alpha\rangle=&(-1)^{n_u+n_h}|\alpha\rangle.
\end{aligned}
\end{equation}
In other words, $|\alpha\rangle$ belongs to the sector $\{Z=(-1)^{n_u+n_h}, X=(-1)^{n_d+n_h}\}$. 

The main result about the relation between the spectrum in model $H$ and model $\tilde{H}$ can be obtained in two steps. Firstly, we need to construct the simultaneous eigenstate of $\mathcal{I}$ and $\tilde{H}$. Recall that $\tilde{H}$ also has bond-centered inversion symmetry, namely, $[\tilde{H},\mathcal{I}]=0$, then, $U_{KT}^{>}|\psi\rangle$ is also an eigenstate of $\tilde{H}$ with the same energy $E$ as $U_{KT}|\psi\rangle$,
\begin{equation}
\begin{aligned}
\tilde{H}U^{>}_{KT}|\psi\rangle=&U_{KT}HU_{KT}\mathcal{I}U_{KT}\mathcal{I}|\psi\rangle\\
=&I\mathcal{I}U_{KT}H|\psi\rangle=IE\mathcal{I}U_{KT}|\psi\rangle\\
=&IEU^{>}_{KT}\mathcal{I}|\psi\rangle=EU_{KT}^{>}|\psi\rangle.
\end{aligned}
\end{equation}
Now, it is easy to construct the simultaneous eigenstates of the $\mathcal{I}$ and $\tilde{H}$, 
\begin{equation}
\begin{aligned}
\mathcal{I}(U_{KT}\pm U^{>}_{KT})|\psi\rangle=&(U_{KT}^{>}\mathcal{I}\pm U_{KT}\mathcal{I})|\psi\rangle\\
=&I(U_{KT}^{>}\pm U_{KT})|\psi\rangle\\
=&\pm I(U_{KT}\pm U^{>}_{KT})|\psi\rangle.
\end{aligned}
\end{equation}
Hence, the eigenstate $|\psi\rangle$ of model $H$ in sector $\{Z,X,I\}$ corresponds to the eigenstate $(U_{KT}+U^{>}_{KT})|\psi\rangle$ in sector $\{Z,X,I\}$ or the eigenstate $(U_{KT}-U^{>}_{KT})|\psi\rangle$ in sector $\{Z,X,-I\}$.

Secondly, we need to know the relation between $U_{KT}|\psi\rangle$ and $U_{KT}^{>}|\psi\rangle$. Since both model $H$ and model $\tilde{H}$ have $Z_2\times Z_2$ symmetry, it is convenient to expand $|\psi\rangle$ into the direct product state $|\alpha\rangle$ in the $Z_2\times Z_2$ basis, i.e. $|\psi\rangle=\sum_{\alpha}c_{\alpha}|\alpha\rangle$. Owing to $|\psi\rangle$ having quantum numbers $Z,X$, the summation is restricted to the states which have the same quantum numbers $Z,X$.

The calculations of $U_{KT}|\alpha\rangle$ and $U_{KT}^{>}|\alpha\rangle$ are straightforward. The transformation $U_{KT}=\prod_{1\leq u<v \leq L}e^{i\pi S^{z}_{u}S^{x}_{v}}$, which is the product of $L(L-1)/2$ terms provided the system size is $L$. We first consider one term in $U_{KT}$, say, $e^{i\pi S^{z}_{i}S^{x}_{j}}$. It is easy to find that
\begin{equation}
\begin{aligned}\label{ukt1}
e^{i\pi S^{z}_{i}S^{x}_{j}}|\uparrow\uparrow\rangle=&|\uparrow\uparrow\rangle,\\
e^{i\pi S^{z}_{i}S^{x}_{j}}|\uparrow h\rangle=&-|\uparrow h\rangle,\\ 
e^{i\pi S^{z}_{i}S^{x}_{j}}|\uparrow\downarrow\rangle=&-|\uparrow\downarrow\rangle,\\
e^{i\pi S^{z}_{i}S^{x}_{j}}|h\uparrow\rangle=&|h\uparrow\rangle,\\
e^{i\pi S^{z}_{i}S^{x}_{j}}|hh\rangle=&-|hh\rangle,\\
e^{i\pi S^{z}_{i}S^{x}_{j}}|h\downarrow\rangle=&-|h\downarrow\rangle,\\
e^{i\pi S^{z}_{i}S^{x}_{j}}|\downarrow\uparrow\rangle=&|\downarrow\uparrow\rangle,\\
e^{i\pi S^{z}_{i}S^{x}_{j}}|\downarrow h\rangle=&|\downarrow h\rangle,\\
e^{i\pi S^{z}_{i}S^{x}_{j}}|\downarrow \downarrow\rangle=&|\downarrow\downarrow\rangle,
\end{aligned}
\end{equation}
where the state $|ab\rangle$ denotes the direct product state of site $i$ and site $j$. These equations mean the direct product state $|\alpha\rangle$ is the eigenstate of $U_{KT}$. Suppose the number of pairs $\uparrow\uparrow$, $\uparrow h$, $\uparrow\downarrow$, $h\uparrow$, $hh$, $h\downarrow$, $\downarrow\uparrow$,$\downarrow h$, $\downarrow\downarrow$ in $|\alpha\rangle$ are $a,b,c,d,e,f,g,h,i$. As there are $L(L-1)/2$ terms in the product of $U_{KT}$, then, we have $a+b+c+\cdots +i=L(L-1)/2$. It is not difficult to show that the number of pairs $\uparrow\uparrow$, $\downarrow\downarrow$, $hh$ in $|\alpha\rangle$ can be written as
\begin{equation}
\begin{aligned}
a=&n_{u}(n_{u}-1)/2,\\ 
i=&n_{d}(n_{d}-1)/2,\\
 e=&n_{h}(n_{h}-1)/2.
\end{aligned}
\end{equation}
Using eq.(\ref{ukt1}), we have 
\begin{equation}\label{uktal}
U_{KT}|\alpha\rangle=(-1)^{b+c+e+f}|\alpha\rangle.
\end{equation}
Similarly, for $U_{KT}^{>}$, we have
\begin{equation}
\begin{aligned}
e^{i\pi S^{x}_{i}S^{z}_{j}}|\uparrow\uparrow\rangle=&|\uparrow\uparrow\rangle,\\
e^{i\pi S^{x}_{i}S^{z}_{j}}|\uparrow h\rangle=&|\uparrow h\rangle,\\ 
e^{i\pi S^{x}_{i}S^{z}_{j}}|\uparrow\downarrow\rangle=&|\uparrow\downarrow\rangle,\\
e^{i\pi S^{x}_{i}S^{z}_{j}}|h\uparrow\rangle=&-|h\uparrow\rangle,\\
e^{i\pi S^{x}_{i}S^{z}_{j}}|hh\rangle=&-|hh\rangle,\\
e^{i\pi S^{x}_{i}S^{z}_{j}}|h\downarrow\rangle=&|h\downarrow\rangle,\\
e^{i\pi S^{x}_{i}S^{z}_{j}}|\downarrow\uparrow\rangle=&-|\downarrow\uparrow\rangle,\\
e^{i\pi S^{x}_{i}S^{z}_{j}}|\downarrow h\rangle=&-|\downarrow h\rangle,\\
e^{i\pi S^{x}_{i}S^{z}_{j}}|\downarrow \downarrow\rangle=&|\downarrow\downarrow\rangle,
\end{aligned}
\end{equation}
and, $|\alpha\rangle$ is also the eigenstate of $U_{KT}^{>}$
\begin{equation}\label{uktlal}
U_{KT}^{>}|\alpha\rangle=(-1)^{d+e+g+h}|\alpha\rangle.
\end{equation}
Therefore, we find that $U_{KT}|\alpha\rangle=\pm U_{KT}^{>}|\alpha\rangle$. To determine the sign, it is convenient to consider $(-1)^{d+e+g+h}\times (-1)^{b+c+e+f}$, which is equivalent to determine the parity of $d+e+g+h+b+c+e+f$. Using above results, we have
\begin{equation}
\begin{aligned}
&b+c+e+f+d+e+g+h\\
=&\frac{L(L-1)}{2}-a-i+e\\
=&\frac{(n_{u}+n_{d}+n_{h})(n_{u}+n_{d}+n_{h}-1)}{2}\\
&-\frac{n_{u}(n_{u}-1)}{2}-\frac{n_{d}(n_{d}-1)}{2}+\frac{n_{h}(n_{h}-1)}{2}\\
=&n_{u}n_{d}+n_{u}n_{h}+n_{d}n_{h}+n_{h}(n_{h}-1),
\end{aligned}
\end{equation}
as $n_{h}(n_{h}-1)$ must be even, the parity of $d+e+g+h+b+c+e+f$ is determined by $n_{u}n_{d}+n_{u}n_{h}+n_{d}n_{h}\equiv sig$. Therefore, $(U_{KT}+U_{KT}^{>})|\alpha\rangle=0$ when $sig$ is odd, $(U_{KT}-U_{KT}^{>})|\alpha\rangle=0$ when $sig$ is even. So far, we find that the quantum numbers $Z,X$ and $sig$ of any direct product state $|\alpha\rangle$ are fully determined by $n_u,n_d,n_h$. And we can classify all possible situations for $n_u,n_d,n_h$, the results are listed in Table.\ref{tab1}. All these situations can be summarized as: 1). for even size $L$, we have $(U_{KT}-U_{KT}^{>})|\alpha\rangle$ when the quantum numbers $Z,X$ are not both 1, otherwise, we have $(U_{KT}+U_{KT}^{>})|\alpha\rangle$. 2). for odd size $L$, we have $(U_{KT}+U_{KT}^{>})|\alpha\rangle$ when the quantum numbers $Z,X$ are not both 1, otherwise, we have $(U_{KT}-U_{KT}^{>})|\alpha\rangle$. Recall that $|\psi\rangle=\sum_{\alpha}c_{\alpha}|\alpha\rangle$ where $|\alpha\rangle$ in the summation has the same quantum numbers $Z,X$ as $|\psi\rangle$, thus it is easy to see that only one of the constructed eigenstate of $\mathcal{I}$ survives when considering $Z,X$. This leads to the relations between the spectrum in model $H$ and model $\tilde{H}$: 
\begin{itemize}
\item For even system size $L$, the eigenstate $|\psi\rangle$ of model $H$ in the sector $\{Z,X,I\}$ corresponds to the eigenstate $(U_{KT}-U_{KT}^{>})|\psi\rangle$ of model $\tilde{H}$ in the sector $\{Z,X,-I\}$ if $Z,X$ are not both 1, otherwise, corresponds to the eigenstate $(U_{KT}+U_{KT}^{>})|\psi\rangle$ of model $\tilde{H}$ in the sector $\{Z,X,I\}$. Accordingly, the levels in the sector $\{Z,X,I\}$ of model $H$ maps to the levels in the sector $\{Z,X,-I\}$ of model $\tilde{H}$ under the Kennedy-Tasaki transformation if $Z,X$ are not both 1, otherwise, maps to the sector $\{Z,X,I\}$ of model $\tilde{H}$.
\item For odd system size $L$, the levels in the sector $\{Z,X,I\}$ of model $H$ maps to the levels in the sector $\{Z,X,I\}$ of model $\tilde{H}$ under the transformation if $Z,X$ are not both 1, otherwise, maps to the sector $\{Z,X,-I\}$ of model $\tilde{H}$.
\end{itemize}
These relations explain the results in Table.\ref{tabsum} nicely. The assumptions we have used are 1). model $H$ has $Z_2\times Z_2$ symmetry and bond-centered inversion symmetry, 2). model $\tilde{H}=U_{KT}HU_{KT}$ also has bond-centered inversion symmetry. Hence, the above relation can be used in other one-dimensional spin-1 models which satisfy the symmetry requirement.

\begin{table}
\caption{All possible situations for the quantum numbers $Z,X$ and $sig$ of a direct product state $|\alpha\rangle$ in the $Z_2\times Z_2$ basis.}
\begin{ruledtabular}
\label{tab1}
\begin{tabular}{cccc|ccc}
$L$ & $n_u$ & $n_d$ & $n_h$ & $Z$ & $X$ & $sig$\\
\hline
even & odd & odd & even  & $-1$ & $-1$ & odd\\
even & odd & even & odd & $1$ & $-1$ & odd\\
even & even & odd & odd  & $-1$ & $1$ &  odd\\
even & even & even & even & $1$ & $1$ &  even\\ 
\hline
odd & odd & odd & odd &  $1$ & $1$  & odd \\
odd & odd & even & even & $-1$ & $1$ & even\\
odd & even & odd & even & $1$ & $-1$ & even\\
odd & even & even & odd & $-1$ & $-1$ & even
\end{tabular}
\end{ruledtabular}
\end{table}

\subsection{"Hidden non-local symmetry"}
Recently, it was found that some models may have "hidden" symmetry, and the level spacing statistics of these models will not show random matrix theory statistics if such "hidden" symmetry has not been resolved \cite{odea2024}. As mentioned in previous section, model $\tilde{H}$ has a "hidden non-local symmetry", where $U_{KT}\sum_{i}S^{z}_{i}U_{KT}$ commutes with $\tilde{H}$. Thus, one may wonder whether a quantum number corresponds to $U_{KT}\sum_{i}S^{z}_{i}U_{KT}$ is neglected. For convenience, we define $S^{z}_{KT}=U_{KT}\sum_{i}S^{z}_{i}U_{KT}$. One can check the relation $U^{x}_{\pi}S^{z}_{KT}=-S^{z}_{KT}U^{x}_{\pi}$ holds, which indicates that the simultaneous eigenstate of $U^{z}_\pi$, $U^{x}_{\pi}$ and $S^{z}_{KT}$ satisfies $S^{z}_{KT}|\psi\rangle=0$. Actually, it is suffices to consider $\sum_{i}S^{z}_i U_{KT}|\psi\rangle=0$. And, we can expand $U_{KT}|\psi\rangle$ in the direct product states of $S^{z}$ basis as $\sum_{\beta}c_{\beta}|\phi_\beta\rangle$, where $\sum_{i}S^{z}_{i}|\phi_\beta\rangle=M_{\beta}|\phi_\beta\rangle$. Then, the condition $\sum_{i}S^{z}_i U_{KT}|\psi\rangle=0$ implies the summation over $\beta$ is restricted to $M^{z}=0$ sector, since $c_\beta$ are not all zero and $|\phi_\beta\rangle$ are linear independent. In other words, states which satisfy $S^{z}_{KT}|\psi\rangle=0$ have quantum number $Z=1$. In short, the quantum number from the "hidden non-local symmetry" $U_{KT}\sum_{i}S^{z}_{i}U_{KT}$ is only compatible with $\{Z=1,X\}$ sector but not $\{Z=-1,X\}$ sector.

Denote the eigenvalue of $S^{z}_{KT}$ as $M^{z}_{KT}$. There are some sectors of model $\tilde{H}$ which are labeled by quantum numbers $\{Z=1,X,I,M^{z}_{KT}=0\}$. For example, consider the eigenstate $|\phi\rangle$ in the $\{M^z=0,X,I\}$ sector of model $H$ where $\sum_{j}S^z_j|\phi\rangle=0$. The corresponding eigenstate $U_{KT}|\phi\rangle$ of model $\tilde{H}$ satisfies $S^{z}_{KT}\,U_{KT}|\phi\rangle=0$. And these states form a sector $\{Z=1,X,I', M^{z}_{KT}=0\}$, since $S^{z}_{KT}\tilde{H}U_{KT}|\phi\rangle=0$. The levels of this sector are identical to those in the $\{M^z=0,X,I\}$ sector of model $H$ and show GOE behavior, as illustrated in Fig.\ref{fig1} (a)(b). However, there is no quantum number $M^{z}_{KT}$ for $Z=-1$ sector. To the best of our knowledge, there is no additional symmetry in this sector, hence, we cannot find GOE behavior in this sector. In contrast to the usual models, where different sectors show roughly the same level spacing statistics \cite{Hsu1993prb,Poilblanc1993,Marco2022prr,Kollath2024}. \emph{Here, the level spacing statistics of the model $\tilde{H}$ exhibit GOE statistics only in $\{Z=1,X,I,M^{z}_{KT}=0\}$ sectors, and shows non-GOE statistics in other sectors labeled by quantum numbers corresponding to the apparent local symmetries $Z,X,I$}. We expect that a similar phenomenon can be found in other models with certain designed unitary transformations if the apparent local symmetries of the model and those of the transformed model are different. As mentioned above, this non-GOE behavior may evolves to Poisson statistics in the thermodynamic limit.

From the above analysis, we observe that the use of the four quantum numbers $Z=1,X,I,M^{z}_{KT}=0$ is essential in model $\tilde{H}$. However, quantum number $Z=1$ becomes redundant in sector $\{M^{z}=0,X,I\}$ of model $H$. 

In this work, model $\tilde{H}$ is regarded as a physical model and this is the first point of view. Then, the apparent local symmetries are identified and utilized in the level spacing statistics. Based on these, we consider the effect of the "hidden non-local symmetry" (this is why we used the term "hidden", as it is relative to the apparent symmetries), and we also distinguish between the local symmetry and non-local symmetry. We find that some sectors exhibit GOE statistics, while the remaining sectors labeled by quantum numbers of apparent local symmetries show non-GOE statistics. Here, we may assert that the level spacing statistics of the model $H$ and model $\tilde{H}$ are consistent from the first point of view, since they both have sectors which show the GOE statistics. 

However, there is a second point of view, where the model $\tilde {H}$ is regarded as a transformed model from model $H$. In this case, one may study the (non-local) transformed symmetry $U_{KT}\sum_{i}S^{z}_{i}U_{KT}$ first, even though it may be difficult to implement in numerical diagonalization. In this framework, the concept of hidden symmetry is no longer applicable. Here, sectors are labeled by $M^{z}_{KT},I$ for $M^{z}_{KT}\neq 0$ and by $M^{z}_{KT}=0,X,I$ for $M^{z}_{KT}=0$. Every sector in model $H$ has a corresponding sector in model $\tilde{H}$, hence, one may expect that the sectors with sufficiently large dimensions will all exhibit GOE statistics. It is clear that the sectors $\{Z=1,X,I,M^{z}_{KT}=0\}$ from the first point of view are the same as the sectors $\{M^{z}_{KT}=0,X,I\}$ from the second point of view. From the second point of view, it is trivial to conclude that unitary equivalent models exhibit the same level spacing statistics in corresponding sectors.

\begin{figure}
	\centering
	\includegraphics[width=0.5\textwidth]{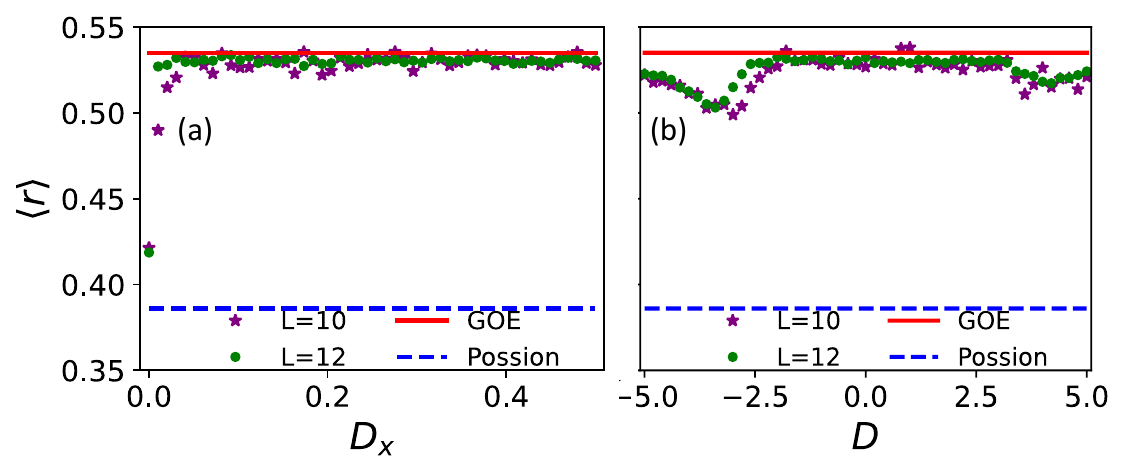}
	\caption{\label{fig3} Mean gap ratio $\langle r\rangle$ as a function of (a) perturbation parameter $D_x$ with fixed $D=1.2$, and (b) single-ion anisotropy $D$ with fixed $D_x=0.1$ for model $\tilde{H}$ in sector $\{Z=-1,X=1,I=1\}$. The red (solid) line and the blue (dashed) line represent the prediction from GOE and Poisson statistics, respectively.}
\end{figure}

\subsection{Perturbations}
The apparent symmetries of model $\tilde{H}$ are $Z_2\times Z_2$ symmetry and bond-centered inversion, we now consider the level spacing statistics under perturbations which preserve the $Z_2\times Z_2$ symmetry while removing the "hidden non-local symmetry" $S^{z}_{KT}$. For the sake of simplicity, we consider the $\{Z=-1,X=1,I=1\}$ sector which does not have a quantum number corresponding to the "hidden symmetry". This sector has dimensions 7381 and 66430 for the system size $L=10, 12$, respectively. The first perturbation is $D_x\sum_{i=1}^{L}(S^{x}_i)^2$, which is invariant under the Kennedy-Tasaki transformation. Once the "hidden symmetry" is removed by small perturbation $D_x$ term (see Fig.\ref{fig3} (a)), the spectrum of a $\{Z,X\}$ sector no longer contains levels from different symmetry sectors,  and the mean gap ratio moves toward the GOE value. With increasing system size, the mean gap ratio reaches GOE value with smaller parameter $D_x$, one would expect that only infinitesimal perturbation is required in the thermodynamic limit. In Fig.\ref{fig3} (b), the mean gap ratio of $\{Z=-1,X=1,I=1\}$ sector with fixed $D_x$ is shown and the result is similar to the usual chaotic model as in Fig.\ref{fig1} (a). 

\begin{figure}
	\centering
	\includegraphics[width=0.5\textwidth]{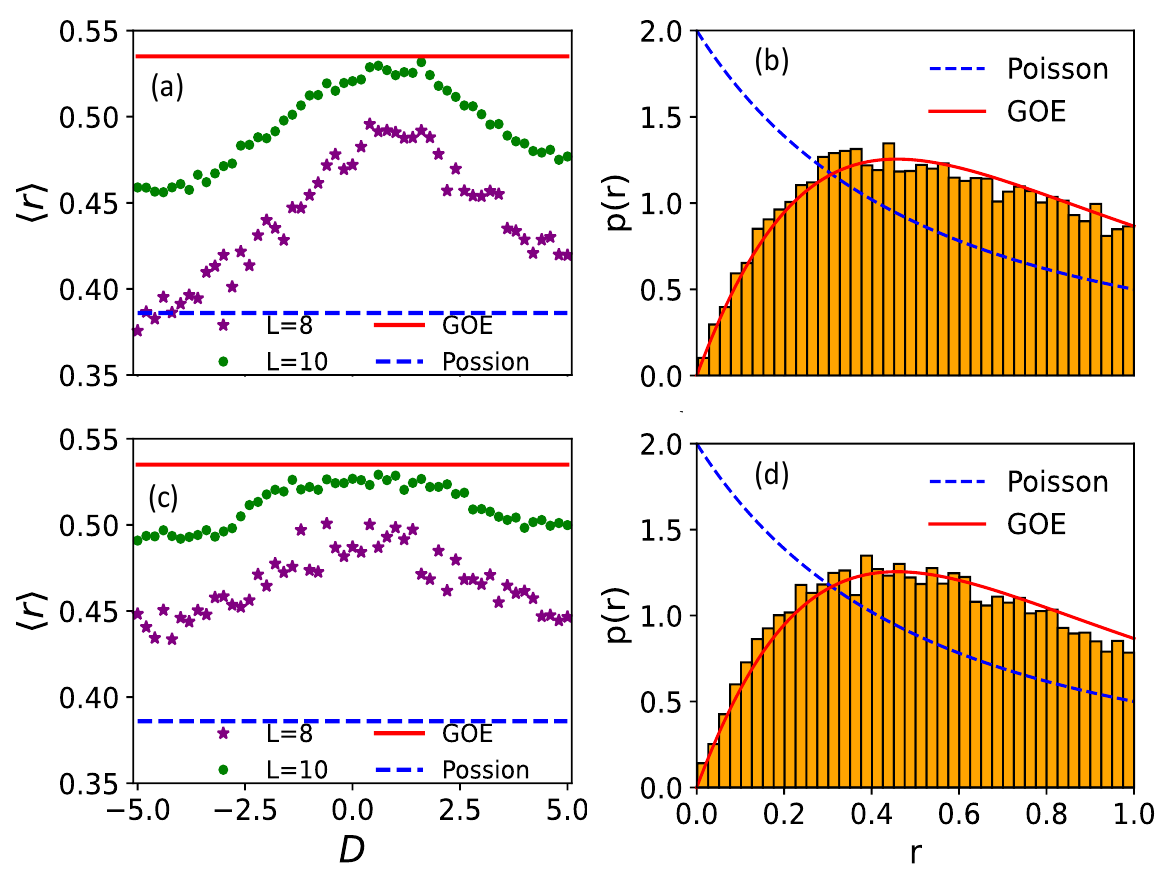}
	\caption{\label{fig4} Mean gap ratio $\langle r\rangle$ as a function of single-ion anisotropy $D$ (a) with edge defect $D_x(S^{x}_1)^2$, (c) with bulk defect $D_x(S^{x}_i)^2$ ($i=L/2$) for model $\tilde{H}$ in sector $\{Z=-1,X=1\}$. Here, $D_x=0.1$. Fixed $D=1.2$ and for the system size $L=10$, gap ratio distribution $P(r)$ with (b) edge defect, (d) bulk defect in the same sector as in (a) (c). The red (solid) line and the blue (dashed) line represent the prediction from GOE and Poisson statistics, respectively.}
\end{figure}

The second perturbation is introducing a defect at a single site. In Fig.\ref{fig4} (a) (b), we consider a single defect $D_x(S^{x}_1)^2$ located at the edge of the chain. In Fig.\ref{fig4} (c) (d), the defect $D_x(S^{x}_i)^2$ is put in the bulk and close to the center of the chain ($i=L/2$). Both these perturbations break bond-centered inversion symmetry, the dimensions of the sector $\{Z=-1,X=1\}$ become 1640 and 14762 for system size $L=8, 10$, respectively. These two kinds of defects are widely used in the spin-1/2 chain \cite{Santos2004,Santos2020prr}. Here, they both modify the level spacing statistics significantly and lead to the GOE statistics even when the strength is relatively weak. 

Suppose we know the conserved quantity in the form as $S^{z}_{KT}=exp(i\pi\sum_{k=1}^{j-1}S^{z}_{k})S_{j}^{z}$, and without the prior knowledge of the transformation and the original model $H$. The results in Fig.\ref{fig3} and Fig.\ref{fig4} may suggest a relation between the non-GOE behavior observed in the $\{Z=-1,X,I\}$ sector and the conserved quantity $S^{z}_{KT}$. It would be interesting to know whether one could find this exact relation by such "measurement" on the model $\tilde{H}$.
%%%%%%%%%%%%%%%%%%%%%%%%%%%%%%%%%%%%%
\section{Conclusions}\label{sectcon}
In this work, we mainly study the spectrum of a one-dimensional spin-1 model $\tilde{H}$ by level spacing statistics. The model $\tilde{H}$ is unitary equivalent to the spin-1 XXZ chain with single-ion anisotropy (model $H$) under the Kennedy-Tasaki transformation. Our main findings are as follows:

i). Two unitary equivalent models can exhibit different level spacing statistics since their apparent local symmetries can be different. Here, the level spacing statistics of the spin-1 XXZ chain follow GOE prediction nicely. While the model $\tilde{H}$ shows non-GOE behavior in its sectors labeled by apparent symmetries.

ii). A chaotic model can display different level spacing statistics in its different symmetry sectors. Here, model $\tilde{H}$ shows GOE behavior in $\{Z=1,X,I,M^{z}_{KT}=0\}$ sectors once the "hidden symmetry" $S^{z}_{KT}$ is resolved. However, other sectors labeled by quantum numbers corresponding to the local symmetries show non-GOE statistics. We may still conclude that the level spacing statistics of the model $H$ and model $\tilde{H}$ are consistent since they both have sectors that show the GOE statistics. 

iii). For the one-dimensional spin-1 model $H$ with $Z_{2}\times Z_{2}$ symmetry and bond-centered inversion symmetry, if its Kennedy-Tasaki transformed model $U_{KT}HU_{KT}$ also has bond-centered inversion symmetry, then the spectrum of sector $\{Z,X,I\}$ of model $H$ will be mapped to the spectrum of sector $\{Z,X,I\}$ or $\{Z,X,-I\}$ of model $U_{KT}HU_{KT}$ under the transformation. The mapping depends on the parity of the system size and $Z,X$.

The non-GOE behavior in model $\tilde{H}$ originates from the spectrum containing levels from different sectors. It is known that the correlation hole in some dynamical quantities, such as survival probability, \cite{Santos2020prr,Hirsch2020pre,Santos2018prb,Santos2019prb} persists even when some symmetries of the model are neglected \cite{Santos2020prr} or when some sectors are mixed \cite{Hirsch2020pre}. The correlation hole can be regarded as a fingerprint of the correlations among the levels \cite{Guhr1998}. It is a dynamical manifestation of the level repulsion \cite{Santos2017}. It would be interesting to study the model with "hidden symmetry" by such dynamical quantities in future research.

%%%%%%%%%%%%%%%%%%%%%%%%%%%%%%%%%%%
\begin{acknowledgements}
We thank Linhao Li, Han-Qing Wu, Haruki Watanabe, Zongping Gong and Hosho Katsura for their helpful discussions. This work is supported by the start-up funding of CQNU (Grant No.24XLB010), the International Postdoctoral Exchange Fellowship Program 2022 by the Office of China Postdoctoral Council: No.PC2022072 and the National Natural Science Foundation of China: No.12147172.
\end{acknowledgements}

\bibliographystyle{apsrev4-1}
\bibliography{mixing_haldane_chaos}

\begin{thebibliography}{46}%
\makeatletter
\providecommand \@ifxundefined [1]{%
 \@ifx{#1\undefined}
}%
\providecommand \@ifnum [1]{%
 \ifnum #1\expandafter \@firstoftwo
 \else \expandafter \@secondoftwo
 \fi
}%
\providecommand \@ifx [1]{%
 \ifx #1\expandafter \@firstoftwo
 \else \expandafter \@secondoftwo
 \fi
}%
\providecommand \natexlab [1]{#1}%
\providecommand \enquote  [1]{``#1''}%
\providecommand \bibnamefont  [1]{#1}%
\providecommand \bibfnamefont [1]{#1}%
\providecommand \citenamefont [1]{#1}%
\providecommand \href@noop [0]{\@secondoftwo}%
\providecommand \href [0]{\begingroup \@sanitize@url \@href}%
\providecommand \@href[1]{\@@startlink{#1}\@@href}%
\providecommand \@@href[1]{\endgroup#1\@@endlink}%
\providecommand \@sanitize@url [0]{\catcode `\\12\catcode `\$12\catcode
  `\&12\catcode `\#12\catcode `\^12\catcode `\_12\catcode `\%12\relax}%
\providecommand \@@startlink[1]{}%
\providecommand \@@endlink[0]{}%
\providecommand \url  [0]{\begingroup\@sanitize@url \@url }%
\providecommand \@url [1]{\endgroup\@href {#1}{\urlprefix }}%
\providecommand \urlprefix  [0]{URL }%
\providecommand \Eprint [0]{\href }%
\providecommand \doibase [0]{http://dx.doi.org/}%
\providecommand \selectlanguage [0]{\@gobble}%
\providecommand \bibinfo  [0]{\@secondoftwo}%
\providecommand \bibfield  [0]{\@secondoftwo}%
\providecommand \translation [1]{[#1]}%
\providecommand \BibitemOpen [0]{}%
\providecommand \bibitemStop [0]{}%
\providecommand \bibitemNoStop [0]{.\EOS\space}%
\providecommand \EOS [0]{\spacefactor3000\relax}%
\providecommand \BibitemShut  [1]{\csname bibitem#1\endcsname}%
\let\auto@bib@innerbib\@empty
%</preamble>
\bibitem [{\citenamefont {Mehta}(2004)}]{mehta2004}%
  \BibitemOpen
  \bibfield  {author} {\bibinfo {author} {\bibfnamefont {M.~L.}\ \bibnamefont
  {Mehta}},\ }\href@noop {} {\emph {\bibinfo {title} {Random matrices}}}\
  (\bibinfo  {publisher} {Elsevier},\ \bibinfo {year} {2004})\BibitemShut
  {NoStop}%
\bibitem [{\citenamefont {Brody}\ \emph {et~al.}(1981)\citenamefont {Brody},
  \citenamefont {Flores}, \citenamefont {French}, \citenamefont {Mello},
  \citenamefont {Pandey},\ and\ \citenamefont {Wong}}]{Brody1981rmp}%
  \BibitemOpen
  \bibfield  {author} {\bibinfo {author} {\bibfnamefont {T.~A.}\ \bibnamefont
  {Brody}}, \bibinfo {author} {\bibfnamefont {J.}~\bibnamefont {Flores}},
  \bibinfo {author} {\bibfnamefont {J.~B.}\ \bibnamefont {French}}, \bibinfo
  {author} {\bibfnamefont {P.~A.}\ \bibnamefont {Mello}}, \bibinfo {author}
  {\bibfnamefont {A.}~\bibnamefont {Pandey}}, \ and\ \bibinfo {author}
  {\bibfnamefont {S.~S.~M.}\ \bibnamefont {Wong}},\ }\href {\doibase
  10.1103/RevModPhys.53.385} {\bibfield  {journal} {\bibinfo  {journal} {Rev.
  Mod. Phys.}\ }\textbf {\bibinfo {volume} {53}},\ \bibinfo {pages} {385}
  (\bibinfo {year} {1981})}\BibitemShut {NoStop}%
\bibitem [{\citenamefont {Edelman}\ \emph {et~al.}(2018)\citenamefont
  {Edelman}, \citenamefont {Macau},\ and\ \citenamefont
  {Sanjuan}}]{edelman2018chaotic}%
  \BibitemOpen
  \bibfield  {author} {\bibinfo {author} {\bibfnamefont {M.}~\bibnamefont
  {Edelman}}, \bibinfo {author} {\bibfnamefont {E.~E.}\ \bibnamefont {Macau}},
  \ and\ \bibinfo {author} {\bibfnamefont {M.~A.}\ \bibnamefont {Sanjuan}},\
  }\href@noop {} {\emph {\bibinfo {title} {Chaotic, fractional, and complex
  dynamics: new insights and perspectives}}}\ (\bibinfo  {publisher}
  {Springer},\ \bibinfo {year} {2018})\BibitemShut {NoStop}%
\bibitem [{\citenamefont {Guhr}\ \emph {et~al.}(1998)\citenamefont {Guhr},
  \citenamefont {Müller–Groeling},\ and\ \citenamefont
  {Weidenmüller}}]{Guhr1998}%
  \BibitemOpen
  \bibfield  {author} {\bibinfo {author} {\bibfnamefont {T.}~\bibnamefont
  {Guhr}}, \bibinfo {author} {\bibfnamefont {A.}~\bibnamefont
  {Müller–Groeling}}, \ and\ \bibinfo {author} {\bibfnamefont {H.~A.}\
  \bibnamefont {Weidenmüller}},\ }\href {\doibase
  https://doi.org/10.1016/S0370-1573(97)00088-4} {\bibfield  {journal}
  {\bibinfo  {journal} {Physics Reports}\ }\textbf {\bibinfo {volume} {299}},\
  \bibinfo {pages} {189} (\bibinfo {year} {1998})}\BibitemShut {NoStop}%
\bibitem [{\citenamefont {Gubin}\ and\ \citenamefont
  {F.~Santos}(2012)}]{Gubin}%
  \BibitemOpen
  \bibfield  {author} {\bibinfo {author} {\bibfnamefont {A.}~\bibnamefont
  {Gubin}}\ and\ \bibinfo {author} {\bibfnamefont {L.}~\bibnamefont
  {F.~Santos}},\ }\href {\doibase 10.1119/1.3671068} {\bibfield  {journal}
  {\bibinfo  {journal} {American Journal of Physics}\ }\textbf {\bibinfo
  {volume} {80}},\ \bibinfo {pages} {246} (\bibinfo {year} {2012})}\BibitemShut
  {NoStop}%
\bibitem [{\citenamefont {Wimberger}(2014)}]{wimberger2014}%
  \BibitemOpen
  \bibfield  {author} {\bibinfo {author} {\bibfnamefont {S.}~\bibnamefont
  {Wimberger}},\ }\href@noop {} {\emph {\bibinfo {title} {Nonlinear dynamics
  and quantum chaos}}},\ Vol.~\bibinfo {volume} {10}\ (\bibinfo  {publisher}
  {Springer},\ \bibinfo {year} {2014})\BibitemShut {NoStop}%
\bibitem [{\citenamefont {Montambaux}\ \emph {et~al.}(1993)\citenamefont
  {Montambaux}, \citenamefont {Poilblanc}, \citenamefont {Bellissard},\ and\
  \citenamefont {Sire}}]{Montambaux1993prl}%
  \BibitemOpen
  \bibfield  {author} {\bibinfo {author} {\bibfnamefont {G.}~\bibnamefont
  {Montambaux}}, \bibinfo {author} {\bibfnamefont {D.}~\bibnamefont
  {Poilblanc}}, \bibinfo {author} {\bibfnamefont {J.}~\bibnamefont
  {Bellissard}}, \ and\ \bibinfo {author} {\bibfnamefont {C.}~\bibnamefont
  {Sire}},\ }\href {\doibase 10.1103/PhysRevLett.70.497} {\bibfield  {journal}
  {\bibinfo  {journal} {Phys. Rev. Lett.}\ }\textbf {\bibinfo {volume} {70}},\
  \bibinfo {pages} {497} (\bibinfo {year} {1993})}\BibitemShut {NoStop}%
\bibitem [{\citenamefont {Poilblanc}\ \emph {et~al.}(1993)\citenamefont
  {Poilblanc}, \citenamefont {Ziman}, \citenamefont {Bellissard}, \citenamefont
  {Mila},\ and\ \citenamefont {Montambaux}}]{Poilblanc1993}%
  \BibitemOpen
  \bibfield  {author} {\bibinfo {author} {\bibfnamefont {D.}~\bibnamefont
  {Poilblanc}}, \bibinfo {author} {\bibfnamefont {T.}~\bibnamefont {Ziman}},
  \bibinfo {author} {\bibfnamefont {J.}~\bibnamefont {Bellissard}}, \bibinfo
  {author} {\bibfnamefont {F.}~\bibnamefont {Mila}}, \ and\ \bibinfo {author}
  {\bibfnamefont {G.}~\bibnamefont {Montambaux}},\ }\href {\doibase
  10.1209/0295-5075/22/7/010} {\bibfield  {journal} {\bibinfo  {journal}
  {Europhysics Letters}\ }\textbf {\bibinfo {volume} {22}},\ \bibinfo {pages}
  {537} (\bibinfo {year} {1993})}\BibitemShut {NoStop}%
\bibitem [{\citenamefont {Hsu}\ and\ \citenamefont
  {Angle`s~d'Auriac}(1993)}]{Hsu1993prb}%
  \BibitemOpen
  \bibfield  {author} {\bibinfo {author} {\bibfnamefont {T.~C.}\ \bibnamefont
  {Hsu}}\ and\ \bibinfo {author} {\bibfnamefont {J.~C.}\ \bibnamefont
  {Angle`s~d'Auriac}},\ }\href {\doibase 10.1103/PhysRevB.47.14291} {\bibfield
  {journal} {\bibinfo  {journal} {Phys. Rev. B}\ }\textbf {\bibinfo {volume}
  {47}},\ \bibinfo {pages} {14291} (\bibinfo {year} {1993})}\BibitemShut
  {NoStop}%
\bibitem [{\citenamefont {Gómez}\ \emph {et~al.}(2011)\citenamefont {Gómez},
  \citenamefont {Kar}, \citenamefont {Kota}, \citenamefont {Molina},
  \citenamefont {Relaño},\ and\ \citenamefont {Retamosa}}]{Gomez2011}%
  \BibitemOpen
  \bibfield  {author} {\bibinfo {author} {\bibfnamefont {J.}~\bibnamefont
  {Gómez}}, \bibinfo {author} {\bibfnamefont {K.}~\bibnamefont {Kar}},
  \bibinfo {author} {\bibfnamefont {V.}~\bibnamefont {Kota}}, \bibinfo {author}
  {\bibfnamefont {R.}~\bibnamefont {Molina}}, \bibinfo {author} {\bibfnamefont
  {A.}~\bibnamefont {Relaño}}, \ and\ \bibinfo {author} {\bibfnamefont
  {J.}~\bibnamefont {Retamosa}},\ }\href {\doibase
  https://doi.org/10.1016/j.physrep.2010.11.003} {\bibfield  {journal}
  {\bibinfo  {journal} {Physics Reports}\ }\textbf {\bibinfo {volume} {499}},\
  \bibinfo {pages} {103} (\bibinfo {year} {2011})}\BibitemShut {NoStop}%
\bibitem [{\citenamefont {Kollath}\ \emph {et~al.}(2010)\citenamefont
  {Kollath}, \citenamefont {Roux}, \citenamefont {Biroli},\ and\ \citenamefont
  {Läuchli}}]{Kollath2010}%
  \BibitemOpen
  \bibfield  {author} {\bibinfo {author} {\bibfnamefont {C.}~\bibnamefont
  {Kollath}}, \bibinfo {author} {\bibfnamefont {G.}~\bibnamefont {Roux}},
  \bibinfo {author} {\bibfnamefont {G.}~\bibnamefont {Biroli}}, \ and\ \bibinfo
  {author} {\bibfnamefont {A.~M.}\ \bibnamefont {Läuchli}},\ }\href {\doibase
  10.1088/1742-5468/2010/08/P08011} {\bibfield  {journal} {\bibinfo  {journal}
  {Journal of Statistical Mechanics: Theory and Experiment}\ }\textbf {\bibinfo
  {volume} {2010}},\ \bibinfo {pages} {P08011} (\bibinfo {year}
  {2010})}\BibitemShut {NoStop}%
\bibitem [{\citenamefont {De~Marco}\ \emph {et~al.}(2022)\citenamefont
  {De~Marco}, \citenamefont {Tolle}, \citenamefont {Halati}, \citenamefont
  {Sheikhan}, \citenamefont {L\"auchli},\ and\ \citenamefont
  {Kollath}}]{Marco2022prr}%
  \BibitemOpen
  \bibfield  {author} {\bibinfo {author} {\bibfnamefont {J.}~\bibnamefont
  {De~Marco}}, \bibinfo {author} {\bibfnamefont {L.}~\bibnamefont {Tolle}},
  \bibinfo {author} {\bibfnamefont {C.-M.}\ \bibnamefont {Halati}}, \bibinfo
  {author} {\bibfnamefont {A.}~\bibnamefont {Sheikhan}}, \bibinfo {author}
  {\bibfnamefont {A.~M.}\ \bibnamefont {L\"auchli}}, \ and\ \bibinfo {author}
  {\bibfnamefont {C.}~\bibnamefont {Kollath}},\ }\href {\doibase
  10.1103/PhysRevResearch.4.033119} {\bibfield  {journal} {\bibinfo  {journal}
  {Phys. Rev. Res.}\ }\textbf {\bibinfo {volume} {4}},\ \bibinfo {pages}
  {033119} (\bibinfo {year} {2022})}\BibitemShut {NoStop}%
\bibitem [{\citenamefont {Haderlein}\ \emph {et~al.}(2024)\citenamefont
  {Haderlein}, \citenamefont {Luitz}, \citenamefont {Kollath},\ and\
  \citenamefont {Sheikhan}}]{Kollath2024}%
  \BibitemOpen
  \bibfield  {author} {\bibinfo {author} {\bibfnamefont {K.}~\bibnamefont
  {Haderlein}}, \bibinfo {author} {\bibfnamefont {D.~J.}\ \bibnamefont
  {Luitz}}, \bibinfo {author} {\bibfnamefont {C.}~\bibnamefont {Kollath}}, \
  and\ \bibinfo {author} {\bibfnamefont {A.}~\bibnamefont {Sheikhan}},\ }\href
  {\doibase 10.1088/1742-5468/ad5270} {\bibfield  {journal} {\bibinfo
  {journal} {Journal of Statistical Mechanics: Theory and Experiment}\ }\textbf
  {\bibinfo {volume} {2024}},\ \bibinfo {pages} {073101} (\bibinfo {year}
  {2024})}\BibitemShut {NoStop}%
\bibitem [{\citenamefont {Santos}(2004)}]{Santos2004}%
  \BibitemOpen
  \bibfield  {author} {\bibinfo {author} {\bibfnamefont {L.~F.}\ \bibnamefont
  {Santos}},\ }\href {\doibase 10.1088/0305-4470/37/17/004} {\bibfield
  {journal} {\bibinfo  {journal} {Journal of Physics A: Mathematical and
  General}\ }\textbf {\bibinfo {volume} {37}},\ \bibinfo {pages} {4723}
  (\bibinfo {year} {2004})}\BibitemShut {NoStop}%
\bibitem [{\citenamefont {Santos}\ \emph {et~al.}(2020)\citenamefont {Santos},
  \citenamefont {P\'erez-Bernal},\ and\ \citenamefont
  {Torres-Herrera}}]{Santos2020prr}%
  \BibitemOpen
  \bibfield  {author} {\bibinfo {author} {\bibfnamefont {L.~F.}\ \bibnamefont
  {Santos}}, \bibinfo {author} {\bibfnamefont {F.}~\bibnamefont
  {P\'erez-Bernal}}, \ and\ \bibinfo {author} {\bibfnamefont {E.~J.}\
  \bibnamefont {Torres-Herrera}},\ }\href {\doibase
  10.1103/PhysRevResearch.2.043034} {\bibfield  {journal} {\bibinfo  {journal}
  {Phys. Rev. Res.}\ }\textbf {\bibinfo {volume} {2}},\ \bibinfo {pages}
  {043034} (\bibinfo {year} {2020})}\BibitemShut {NoStop}%
\bibitem [{\citenamefont {Sanada}\ \emph {et~al.}(2023)\citenamefont {Sanada},
  \citenamefont {Miao},\ and\ \citenamefont {Katsura}}]{Katsura2023}%
  \BibitemOpen
  \bibfield  {author} {\bibinfo {author} {\bibfnamefont {K.}~\bibnamefont
  {Sanada}}, \bibinfo {author} {\bibfnamefont {Y.}~\bibnamefont {Miao}}, \ and\
  \bibinfo {author} {\bibfnamefont {H.}~\bibnamefont {Katsura}},\ }\href
  {\doibase 10.1103/PhysRevB.108.155102} {\bibfield  {journal} {\bibinfo
  {journal} {Phys. Rev. B}\ }\textbf {\bibinfo {volume} {108}},\ \bibinfo
  {pages} {155102} (\bibinfo {year} {2023})}\BibitemShut {NoStop}%
\bibitem [{\citenamefont {O'Dea}(2024)}]{odea2024}%
  \BibitemOpen
  \bibfield  {author} {\bibinfo {author} {\bibfnamefont {N.}~\bibnamefont
  {O'Dea}},\ }\href@noop {} {\bibfield  {journal} {\bibinfo  {journal} {arXiv
  preprint arXiv:2406.03983}\ } (\bibinfo {year} {2024})}\BibitemShut {NoStop}%
\bibitem [{\citenamefont {Heilmann}\ and\ \citenamefont
  {Lieb}(1971)}]{Lieb1971}%
  \BibitemOpen
  \bibfield  {author} {\bibinfo {author} {\bibfnamefont {O.~J.}\ \bibnamefont
  {Heilmann}}\ and\ \bibinfo {author} {\bibfnamefont {E.~H.}\ \bibnamefont
  {Lieb}},\ }\href {\doibase
  https://doi.org/10.1111/j.1749-6632.1971.tb34956.x} {\bibfield  {journal}
  {\bibinfo  {journal} {Annals of the New York Academy of Sciences}\ }\textbf
  {\bibinfo {volume} {172}},\ \bibinfo {pages} {584} (\bibinfo {year}
  {1971})}\BibitemShut {NoStop}%
\bibitem [{\citenamefont {Kudo}\ and\ \citenamefont
  {Deguchi}(2005)}]{Kudo2005}%
  \BibitemOpen
  \bibfield  {author} {\bibinfo {author} {\bibfnamefont {K.}~\bibnamefont
  {Kudo}}\ and\ \bibinfo {author} {\bibfnamefont {T.}~\bibnamefont {Deguchi}},\
  }\href {\doibase 10.1143/JPSJ.74.1992} {\bibfield  {journal} {\bibinfo
  {journal} {Journal of the Physical Society of Japan}\ }\textbf {\bibinfo
  {volume} {74}},\ \bibinfo {pages} {1992} (\bibinfo {year}
  {2005})}\BibitemShut {NoStop}%
\bibitem [{\citenamefont {Rosenzweig}\ and\ \citenamefont
  {Porter}(1960)}]{Porter1960pr}%
  \BibitemOpen
  \bibfield  {author} {\bibinfo {author} {\bibfnamefont {N.}~\bibnamefont
  {Rosenzweig}}\ and\ \bibinfo {author} {\bibfnamefont {C.~E.}\ \bibnamefont
  {Porter}},\ }\href {\doibase 10.1103/PhysRev.120.1698} {\bibfield  {journal}
  {\bibinfo  {journal} {Phys. Rev.}\ }\textbf {\bibinfo {volume} {120}},\
  \bibinfo {pages} {1698} (\bibinfo {year} {1960})}\BibitemShut {NoStop}%
\bibitem [{\citenamefont {Giraud}\ \emph {et~al.}(2022)\citenamefont {Giraud},
  \citenamefont {Mac\'e}, \citenamefont {Vernier},\ and\ \citenamefont
  {Alet}}]{Alet2022prx}%
  \BibitemOpen
  \bibfield  {author} {\bibinfo {author} {\bibfnamefont {O.}~\bibnamefont
  {Giraud}}, \bibinfo {author} {\bibfnamefont {N.}~\bibnamefont {Mac\'e}},
  \bibinfo {author} {\bibfnamefont {E.}~\bibnamefont {Vernier}}, \ and\
  \bibinfo {author} {\bibfnamefont {F.}~\bibnamefont {Alet}},\ }\href {\doibase
  10.1103/PhysRevX.12.011006} {\bibfield  {journal} {\bibinfo  {journal} {Phys.
  Rev. X}\ }\textbf {\bibinfo {volume} {12}},\ \bibinfo {pages} {011006}
  (\bibinfo {year} {2022})}\BibitemShut {NoStop}%
\bibitem [{\citenamefont {Kennedy}\ and\ \citenamefont
  {Tasaki}(1992{\natexlab{a}})}]{Kennedy1992}%
  \BibitemOpen
  \bibfield  {author} {\bibinfo {author} {\bibfnamefont {T.}~\bibnamefont
  {Kennedy}}\ and\ \bibinfo {author} {\bibfnamefont {H.}~\bibnamefont
  {Tasaki}},\ }\href {\doibase 10.1007/BF02097239} {\bibfield  {journal}
  {\bibinfo  {journal} {Communications in Mathematical Physics}\ }\textbf
  {\bibinfo {volume} {147}},\ \bibinfo {pages} {431} (\bibinfo {year}
  {1992}{\natexlab{a}})}\BibitemShut {NoStop}%
\bibitem [{\citenamefont {Kennedy}\ and\ \citenamefont
  {Tasaki}(1992{\natexlab{b}})}]{Kennedy1992prb}%
  \BibitemOpen
  \bibfield  {author} {\bibinfo {author} {\bibfnamefont {T.}~\bibnamefont
  {Kennedy}}\ and\ \bibinfo {author} {\bibfnamefont {H.}~\bibnamefont
  {Tasaki}},\ }\href {\doibase 10.1103/PhysRevB.45.304} {\bibfield  {journal}
  {\bibinfo  {journal} {Phys. Rev. B}\ }\textbf {\bibinfo {volume} {45}},\
  \bibinfo {pages} {304} (\bibinfo {year} {1992}{\natexlab{b}})}\BibitemShut
  {NoStop}%
\bibitem [{\citenamefont {Oshikawa}(1992)}]{Oshikawa1992}%
  \BibitemOpen
  \bibfield  {author} {\bibinfo {author} {\bibfnamefont {M.}~\bibnamefont
  {Oshikawa}},\ }\href {\doibase 10.1088/0953-8984/4/36/019} {\bibfield
  {journal} {\bibinfo  {journal} {Journal of Physics: Condensed Matter}\
  }\textbf {\bibinfo {volume} {4}},\ \bibinfo {pages} {7469} (\bibinfo {year}
  {1992})}\BibitemShut {NoStop}%
\bibitem [{\citenamefont {Tasaki}(2020)}]{tasaki2020}%
  \BibitemOpen
  \bibfield  {author} {\bibinfo {author} {\bibfnamefont {H.}~\bibnamefont
  {Tasaki}},\ }\href@noop {} {\emph {\bibinfo {title} {Physics and mathematics
  of quantum many-body systems}}},\ Vol.~\bibinfo {volume} {66}\ (\bibinfo
  {publisher} {Springer},\ \bibinfo {year} {2020})\BibitemShut {NoStop}%
\bibitem [{\citenamefont {Haldane}(1983{\natexlab{a}})}]{Haldane1983}%
  \BibitemOpen
  \bibfield  {author} {\bibinfo {author} {\bibfnamefont {F.}~\bibnamefont
  {Haldane}},\ }\href {\doibase https://doi.org/10.1016/0375-9601(83)90631-X}
  {\bibfield  {journal} {\bibinfo  {journal} {Physics Letters A}\ }\textbf
  {\bibinfo {volume} {93}},\ \bibinfo {pages} {464} (\bibinfo {year}
  {1983}{\natexlab{a}})}\BibitemShut {NoStop}%
\bibitem [{\citenamefont {Haldane}(1983{\natexlab{b}})}]{Haldane1983prl}%
  \BibitemOpen
  \bibfield  {author} {\bibinfo {author} {\bibfnamefont {F.~D.~M.}\
  \bibnamefont {Haldane}},\ }\href {\doibase 10.1103/PhysRevLett.50.1153}
  {\bibfield  {journal} {\bibinfo  {journal} {Phys. Rev. Lett.}\ }\textbf
  {\bibinfo {volume} {50}},\ \bibinfo {pages} {1153} (\bibinfo {year}
  {1983}{\natexlab{b}})}\BibitemShut {NoStop}%
\bibitem [{\citenamefont {Gu}\ and\ \citenamefont {Wen}(2009)}]{ZCGu2009prb}%
  \BibitemOpen
  \bibfield  {author} {\bibinfo {author} {\bibfnamefont {Z.-C.}\ \bibnamefont
  {Gu}}\ and\ \bibinfo {author} {\bibfnamefont {X.-G.}\ \bibnamefont {Wen}},\
  }\href {\doibase 10.1103/PhysRevB.80.155131} {\bibfield  {journal} {\bibinfo
  {journal} {Phys. Rev. B}\ }\textbf {\bibinfo {volume} {80}},\ \bibinfo
  {pages} {155131} (\bibinfo {year} {2009})}\BibitemShut {NoStop}%
\bibitem [{\citenamefont {Pollmann}\ \emph {et~al.}(2010)\citenamefont
  {Pollmann}, \citenamefont {Turner}, \citenamefont {Berg},\ and\ \citenamefont
  {Oshikawa}}]{Pollmann2010prb}%
  \BibitemOpen
  \bibfield  {author} {\bibinfo {author} {\bibfnamefont {F.}~\bibnamefont
  {Pollmann}}, \bibinfo {author} {\bibfnamefont {A.~M.}\ \bibnamefont
  {Turner}}, \bibinfo {author} {\bibfnamefont {E.}~\bibnamefont {Berg}}, \ and\
  \bibinfo {author} {\bibfnamefont {M.}~\bibnamefont {Oshikawa}},\ }\href
  {\doibase 10.1103/PhysRevB.81.064439} {\bibfield  {journal} {\bibinfo
  {journal} {Phys. Rev. B}\ }\textbf {\bibinfo {volume} {81}},\ \bibinfo
  {pages} {064439} (\bibinfo {year} {2010})}\BibitemShut {NoStop}%
\bibitem [{\citenamefont {Pollmann}\ \emph {et~al.}(2012)\citenamefont
  {Pollmann}, \citenamefont {Berg}, \citenamefont {Turner},\ and\ \citenamefont
  {Oshikawa}}]{Pollmann2012prb}%
  \BibitemOpen
  \bibfield  {author} {\bibinfo {author} {\bibfnamefont {F.}~\bibnamefont
  {Pollmann}}, \bibinfo {author} {\bibfnamefont {E.}~\bibnamefont {Berg}},
  \bibinfo {author} {\bibfnamefont {A.~M.}\ \bibnamefont {Turner}}, \ and\
  \bibinfo {author} {\bibfnamefont {M.}~\bibnamefont {Oshikawa}},\ }\href
  {\doibase 10.1103/PhysRevB.85.075125} {\bibfield  {journal} {\bibinfo
  {journal} {Phys. Rev. B}\ }\textbf {\bibinfo {volume} {85}},\ \bibinfo
  {pages} {075125} (\bibinfo {year} {2012})}\BibitemShut {NoStop}%
\bibitem [{\citenamefont {Li}\ \emph {et~al.}(2023{\natexlab{a}})\citenamefont
  {Li}, \citenamefont {Oshikawa},\ and\ \citenamefont {Zheng}}]{li2023non}%
  \BibitemOpen
  \bibfield  {author} {\bibinfo {author} {\bibfnamefont {L.}~\bibnamefont
  {Li}}, \bibinfo {author} {\bibfnamefont {M.}~\bibnamefont {Oshikawa}}, \ and\
  \bibinfo {author} {\bibfnamefont {Y.}~\bibnamefont {Zheng}},\ }\href
  {\doibase 10.1103/PhysRevB.108.214429} {\bibfield  {journal} {\bibinfo
  {journal} {Phys. Rev. B}\ }\textbf {\bibinfo {volume} {108}},\ \bibinfo
  {pages} {214429} (\bibinfo {year} {2023}{\natexlab{a}})}\BibitemShut
  {NoStop}%
\bibitem [{\citenamefont {Li}\ \emph {et~al.}(2023{\natexlab{b}})\citenamefont
  {Li}, \citenamefont {Oshikawa},\ and\ \citenamefont
  {Zheng}}]{li2023intrinsically}%
  \BibitemOpen
  \bibfield  {author} {\bibinfo {author} {\bibfnamefont {L.}~\bibnamefont
  {Li}}, \bibinfo {author} {\bibfnamefont {M.}~\bibnamefont {Oshikawa}}, \ and\
  \bibinfo {author} {\bibfnamefont {Y.}~\bibnamefont {Zheng}},\ }\href@noop {}
  {\bibfield  {journal} {\bibinfo  {journal} {arXiv preprint arXiv:2307.04788}\
  } (\bibinfo {year} {2023}{\natexlab{b}})}\BibitemShut {NoStop}%
\bibitem [{\citenamefont {Okunishi}(2011)}]{Okunishi2011prb}%
  \BibitemOpen
  \bibfield  {author} {\bibinfo {author} {\bibfnamefont {K.}~\bibnamefont
  {Okunishi}},\ }\href {\doibase 10.1103/PhysRevB.83.104411} {\bibfield
  {journal} {\bibinfo  {journal} {Phys. Rev. B}\ }\textbf {\bibinfo {volume}
  {83}},\ \bibinfo {pages} {104411} (\bibinfo {year} {2011})}\BibitemShut
  {NoStop}%
\bibitem [{\citenamefont {Chen}\ \emph {et~al.}(2003)\citenamefont {Chen},
  \citenamefont {Hida},\ and\ \citenamefont {Sanctuary}}]{WeiChen2003prb}%
  \BibitemOpen
  \bibfield  {author} {\bibinfo {author} {\bibfnamefont {W.}~\bibnamefont
  {Chen}}, \bibinfo {author} {\bibfnamefont {K.}~\bibnamefont {Hida}}, \ and\
  \bibinfo {author} {\bibfnamefont {B.~C.}\ \bibnamefont {Sanctuary}},\ }\href
  {\doibase 10.1103/PhysRevB.67.104401} {\bibfield  {journal} {\bibinfo
  {journal} {Phys. Rev. B}\ }\textbf {\bibinfo {volume} {67}},\ \bibinfo
  {pages} {104401} (\bibinfo {year} {2003})}\BibitemShut {NoStop}%
\bibitem [{\citenamefont {Weinberg}\ and\ \citenamefont
  {Bukov}(2017)}]{quspin1}%
  \BibitemOpen
  \bibfield  {author} {\bibinfo {author} {\bibfnamefont {P.}~\bibnamefont
  {Weinberg}}\ and\ \bibinfo {author} {\bibfnamefont {M.}~\bibnamefont
  {Bukov}},\ }\href {\doibase 10.21468/SciPostPhys.2.1.003} {\bibfield
  {journal} {\bibinfo  {journal} {SciPost Phys.}\ }\textbf {\bibinfo {volume}
  {2}},\ \bibinfo {pages} {003} (\bibinfo {year} {2017})}\BibitemShut {NoStop}%
\bibitem [{\citenamefont {Weinberg}\ and\ \citenamefont
  {Bukov}(2019)}]{quspin2}%
  \BibitemOpen
  \bibfield  {author} {\bibinfo {author} {\bibfnamefont {P.}~\bibnamefont
  {Weinberg}}\ and\ \bibinfo {author} {\bibfnamefont {M.}~\bibnamefont
  {Bukov}},\ }\href {\doibase 10.21468/SciPostPhys.7.2.020} {\bibfield
  {journal} {\bibinfo  {journal} {SciPost Phys.}\ }\textbf {\bibinfo {volume}
  {7}},\ \bibinfo {pages} {020} (\bibinfo {year} {2019})}\BibitemShut {NoStop}%
\bibitem [{\citenamefont {Kennedy}(1994)}]{Kennedy1994}%
  \BibitemOpen
  \bibfield  {author} {\bibinfo {author} {\bibfnamefont {T.}~\bibnamefont
  {Kennedy}},\ }\href {\doibase 10.1088/0953-8984/6/39/020} {\bibfield
  {journal} {\bibinfo  {journal} {Journal of Physics: Condensed Matter}\
  }\textbf {\bibinfo {volume} {6}},\ \bibinfo {pages} {8015} (\bibinfo {year}
  {1994})}\BibitemShut {NoStop}%
\bibitem [{\citenamefont {Yang}\ \emph {et~al.}(2023)\citenamefont {Yang},
  \citenamefont {Li}, \citenamefont {Okunishi},\ and\ \citenamefont
  {Katsura}}]{HongYang2023prb}%
  \BibitemOpen
  \bibfield  {author} {\bibinfo {author} {\bibfnamefont {H.}~\bibnamefont
  {Yang}}, \bibinfo {author} {\bibfnamefont {L.}~\bibnamefont {Li}}, \bibinfo
  {author} {\bibfnamefont {K.}~\bibnamefont {Okunishi}}, \ and\ \bibinfo
  {author} {\bibfnamefont {H.}~\bibnamefont {Katsura}},\ }\href {\doibase
  10.1103/PhysRevB.107.125158} {\bibfield  {journal} {\bibinfo  {journal}
  {Phys. Rev. B}\ }\textbf {\bibinfo {volume} {107}},\ \bibinfo {pages}
  {125158} (\bibinfo {year} {2023})}\BibitemShut {NoStop}%
\bibitem [{\citenamefont {Kudo}\ and\ \citenamefont
  {Deguchi}(2003)}]{Kudo2003prb}%
  \BibitemOpen
  \bibfield  {author} {\bibinfo {author} {\bibfnamefont {K.}~\bibnamefont
  {Kudo}}\ and\ \bibinfo {author} {\bibfnamefont {T.}~\bibnamefont {Deguchi}},\
  }\href {\doibase 10.1103/PhysRevB.68.052510} {\bibfield  {journal} {\bibinfo
  {journal} {Phys. Rev. B}\ }\textbf {\bibinfo {volume} {68}},\ \bibinfo
  {pages} {052510} (\bibinfo {year} {2003})}\BibitemShut {NoStop}%
\bibitem [{\citenamefont {Oganesyan}\ and\ \citenamefont
  {Huse}(2007)}]{Huse2007prb}%
  \BibitemOpen
  \bibfield  {author} {\bibinfo {author} {\bibfnamefont {V.}~\bibnamefont
  {Oganesyan}}\ and\ \bibinfo {author} {\bibfnamefont {D.~A.}\ \bibnamefont
  {Huse}},\ }\href {\doibase 10.1103/PhysRevB.75.155111} {\bibfield  {journal}
  {\bibinfo  {journal} {Phys. Rev. B}\ }\textbf {\bibinfo {volume} {75}},\
  \bibinfo {pages} {155111} (\bibinfo {year} {2007})}\BibitemShut {NoStop}%
\bibitem [{\citenamefont {Atas}\ \emph {et~al.}(2013)\citenamefont {Atas},
  \citenamefont {Bogomolny}, \citenamefont {Giraud},\ and\ \citenamefont
  {Roux}}]{Atas2013prl}%
  \BibitemOpen
  \bibfield  {author} {\bibinfo {author} {\bibfnamefont {Y.~Y.}\ \bibnamefont
  {Atas}}, \bibinfo {author} {\bibfnamefont {E.}~\bibnamefont {Bogomolny}},
  \bibinfo {author} {\bibfnamefont {O.}~\bibnamefont {Giraud}}, \ and\ \bibinfo
  {author} {\bibfnamefont {G.}~\bibnamefont {Roux}},\ }\href {\doibase
  10.1103/PhysRevLett.110.084101} {\bibfield  {journal} {\bibinfo  {journal}
  {Phys. Rev. Lett.}\ }\textbf {\bibinfo {volume} {110}},\ \bibinfo {pages}
  {084101} (\bibinfo {year} {2013})}\BibitemShut {NoStop}%
\bibitem [{\citenamefont {Moudgalya}\ \emph {et~al.}(2018)\citenamefont
  {Moudgalya}, \citenamefont {Rachel}, \citenamefont {Bernevig},\ and\
  \citenamefont {Regnault}}]{Sanjay2018prb}%
  \BibitemOpen
  \bibfield  {author} {\bibinfo {author} {\bibfnamefont {S.}~\bibnamefont
  {Moudgalya}}, \bibinfo {author} {\bibfnamefont {S.}~\bibnamefont {Rachel}},
  \bibinfo {author} {\bibfnamefont {B.~A.}\ \bibnamefont {Bernevig}}, \ and\
  \bibinfo {author} {\bibfnamefont {N.}~\bibnamefont {Regnault}},\ }\href
  {\doibase 10.1103/PhysRevB.98.235155} {\bibfield  {journal} {\bibinfo
  {journal} {Phys. Rev. B}\ }\textbf {\bibinfo {volume} {98}},\ \bibinfo
  {pages} {235155} (\bibinfo {year} {2018})}\BibitemShut {NoStop}%
\bibitem [{\citenamefont {de~la Cruz}\ \emph {et~al.}(2020)\citenamefont {de~la
  Cruz}, \citenamefont {Lerma-Hern\'andez},\ and\ \citenamefont
  {Hirsch}}]{Hirsch2020pre}%
  \BibitemOpen
  \bibfield  {author} {\bibinfo {author} {\bibfnamefont {J.}~\bibnamefont
  {de~la Cruz}}, \bibinfo {author} {\bibfnamefont {S.}~\bibnamefont
  {Lerma-Hern\'andez}}, \ and\ \bibinfo {author} {\bibfnamefont {J.~G.}\
  \bibnamefont {Hirsch}},\ }\href {\doibase 10.1103/PhysRevE.102.032208}
  {\bibfield  {journal} {\bibinfo  {journal} {Phys. Rev. E}\ }\textbf {\bibinfo
  {volume} {102}},\ \bibinfo {pages} {032208} (\bibinfo {year}
  {2020})}\BibitemShut {NoStop}%
\bibitem [{\citenamefont {Torres-Herrera}\ \emph {et~al.}(2018)\citenamefont
  {Torres-Herrera}, \citenamefont {Garc\'{\i}a-Garc\'{\i}a},\ and\
  \citenamefont {Santos}}]{Santos2018prb}%
  \BibitemOpen
  \bibfield  {author} {\bibinfo {author} {\bibfnamefont {E.~J.}\ \bibnamefont
  {Torres-Herrera}}, \bibinfo {author} {\bibfnamefont {A.~M.}\ \bibnamefont
  {Garc\'{\i}a-Garc\'{\i}a}}, \ and\ \bibinfo {author} {\bibfnamefont {L.~F.}\
  \bibnamefont {Santos}},\ }\href {\doibase 10.1103/PhysRevB.97.060303}
  {\bibfield  {journal} {\bibinfo  {journal} {Phys. Rev. B}\ }\textbf {\bibinfo
  {volume} {97}},\ \bibinfo {pages} {060303} (\bibinfo {year}
  {2018})}\BibitemShut {NoStop}%
\bibitem [{\citenamefont {Schiulaz}\ \emph {et~al.}(2019)\citenamefont
  {Schiulaz}, \citenamefont {Torres-Herrera},\ and\ \citenamefont
  {Santos}}]{Santos2019prb}%
  \BibitemOpen
  \bibfield  {author} {\bibinfo {author} {\bibfnamefont {M.}~\bibnamefont
  {Schiulaz}}, \bibinfo {author} {\bibfnamefont {E.~J.}\ \bibnamefont
  {Torres-Herrera}}, \ and\ \bibinfo {author} {\bibfnamefont {L.~F.}\
  \bibnamefont {Santos}},\ }\href {\doibase 10.1103/PhysRevB.99.174313}
  {\bibfield  {journal} {\bibinfo  {journal} {Phys. Rev. B}\ }\textbf {\bibinfo
  {volume} {99}},\ \bibinfo {pages} {174313} (\bibinfo {year}
  {2019})}\BibitemShut {NoStop}%
\bibitem [{\citenamefont {Torres-Herrera}\ and\ \citenamefont
  {Santos}(2017)}]{Santos2017}%
  \BibitemOpen
  \bibfield  {author} {\bibinfo {author} {\bibfnamefont {E.~J.}\ \bibnamefont
  {Torres-Herrera}}\ and\ \bibinfo {author} {\bibfnamefont {L.~F.}\
  \bibnamefont {Santos}},\ }\href {\doibase 10.1098/rsta.2016.0434} {\bibfield
  {journal} {\bibinfo  {journal} {Philosophical Transactions of the Royal
  Society A: Mathematical, Physical and Engineering Sciences}\ }\textbf
  {\bibinfo {volume} {375}},\ \bibinfo {pages} {20160434} (\bibinfo {year}
  {2017})}\BibitemShut {NoStop}%
\end{thebibliography}%
\end{document}